%% file: arxiv_submission_2.tex
\documentclass[reprint,amsmath,amssymb,aps,prb,superscriptaddress]{revtex4-2}

\usepackage[dvipdfmx]{graphicx}
\usepackage{dcolumn}
\usepackage{bm}
\usepackage{color}
\usepackage{physics}
\usepackage{comment}
\begin{document}
\newcommand{\DeltaMs}{$\Delta M_s=\pm 2$ }
\newcommand{\DeltaMsone}{$\Delta M_s=\pm 1$ }

\preprint{APS/123-QED}

\title{Overtone Rabi oscillation of optically polarized triplet electron spins and nuclear hyperpolarization in powder}

\author{Koichiro Miyanishi}
\email{miyanishi.koichiro.qiqb@osaka-u.ac.jp; Present address: Qubitcore Inc., OIST Inovation core2 OIC2 207 1919-1 Tancha, Onna-son, Kunigami-gun Okinawa 904-0495, Japan}
\affiliation{Center for Quantum Information and Quantum Biology, The University of Osaka, 1-2 Machikaneyama, Toyonaka, Osaka 560-0043 Japan}

\author{Takuya F. Segawa}
\affiliation{Institute of Molecular Physical Science / Laboratory of Physical Chemistry, Department of Chemistry and Applied Biosciences, ETH Zurich, 8093 Zurich, Switzerland}

\author{Makoto Negoro}
\affiliation{Center for Quantum Information and Quantum Biology, The University of Osaka, 1-2 Machikaneyama, Toyonaka, Osaka 560-0043 Japan}
\affiliation{Institute for Quantum Life Science (iQLS), National Institutes for Quantum Science and Technology (QST), Chiba 263-8555, Japan}
\affiliation{Premium Research Institute for Human Metaverse Medicine, The University of Osaka, Suita, Osaka 565-0871, Japan}

\author{Akinori Kagawa}
\affiliation{Center for Quantum Information and Quantum Biology, The University of Osaka, 1-2 Machikaneyama, Toyonaka, Osaka 560-0043 Japan}
\affiliation{Premium Research Institute for Human Metaverse Medicine, The University of Osaka, Suita, Osaka 565-0871, Japan}

\author{Kazuyuki Takeda}
\email{takezo@kuchem.kyoto-u.ac.jp}
\affiliation{ Division of Chemistry, Graduate School of Science, Kyoto University, Kyoto 606-8502, Japan}

\date{\today}
\begin{abstract}
We report coherent overtone Rabi oscillations of optically-polarized triplet electron spins and nuclear hyperpolarization in powder samples at room temperature. The strong dependence of the single-quantum resonance on the orientation of the zero-field splitting (ZFS) interaction is overcome by coherently driving the significantly narrower overtone transition. Analytical formulas for the overtone lineshape and nutation functions for axially symmetric ZFS interactions are derived. Overtone Rabi oscillations are observed in pentacene-doped \textit{p}-terphenyl and NV$^-$ centers in microdiamonds.  For the former, overtone triplet dynamic nuclear polarization using the integrated solid effect leads to $^1$H spin polarization of $0.183\pm0.005$\% at a magnetic field of 0.2~T. The $^1$H NMR polarization is enhanced by a factor of 2600 with respect to thermal equilibrium and reaches a large portion of the randomly oriented microcrystals.
\end{abstract}

\maketitle

\section{Introduction}

Nuclear magnetic resonance~(NMR) and magnetic resonance imaging~(MRI) enable chemical analysis and anatomical diagnosis through nuclear induction signals and are utilized in various fields of science, technology and medicine.
NMR and MRI measurements in combination with dynamic nuclear polarization~(DNP)~\cite{overhauser1953polarization} have gained growing popularity due to the significantly boosted sensitivity, and are applied in various circumstances.
Recent remarkable examples include DNP combined with magic angle spinning (MAS) for solid-state NMR~\cite{barnes2010resolution, Lesage10} and dissolution-DNP for liquid-state NMR spectroscopy and MRI~\cite{Ardenkjaer03,Nelson13}, where electron spin polarization of free radicals is fully exploited.

With DNP using free radicals, both cryogenic temperature ($<20$~K) and a relatively high magnetic field ($>3$~T) are required to enhance proton ($^{1}$H) polarization over ca.~660 times (the ratio of the electron's gyromagnetic ratio to the proton's) compared to the thermal $^{1}$H polarization at room temperature.
Alternatively, optically polarized electron spins in the triplet state can lead to nuclear hyperpolarization even at ambient temperatures and in relatively low magnetic fields.
Two representative examples of the source of spin polarization in such \textit{triplet-DNP} schemes are the photo-excited triplet state of pentacene~\cite{henstra1990high,Hautle2024} and negatively charged nitrogen-vacancy (NV$^{-}$) color centers in diamond~\cite{parkerOpticallyPumpedDynamic2019}.
Indeed, $^{1}$H spins in a single crystal of pentacene-doped \textit{p}-terphenyl and $^{13}$C spins in a diamond crystal containing NV$^{-}$ centers were polarized to 34~\%~\cite{Tateishi14} and 6~\%~\cite{King15RoomTemperatureDNP6percent} at room temperature.
When the sources of electron polarization are in the photo-excited triplet state, the absence of paramagnetic electrons during the intervals between repeatedly applied pulsed photo-excitation removes barriers of $^{1}$H spin diffusion, effectively fostering the buildup of $^{1}$H polarization.
Other approaches to nuclear hyperpolarization methods that do not rely on unpaired electrons include parahydrogen-induced polarization~(PHIP), photo-chemically induced DNP~(photo-CIDNP) and spin exchange optical pumping~(SEOP)~\cite{eillsSpinHyperpolarizationModern2023}. 

Triplet-DNP was also demonstrated in polycrystalline samples~\cite{takeda2001dynamic} and successfully applied to dissolution DNP~\cite{negoro2018dissolution,KAGAWA2019dissolution_tripletDNP}.
However, the nuclear polarization achieved in triplet-DNP in powder remains much lower than in a single crystal.
This is due to the anisotropy of the zero-field splitting (ZFS) interaction, which significantly broadens the electron paramagnetic resonance (EPR) of the triplet state, limiting the DNP enhancement to only a tiny fraction of the crystallites and leaving nuclei in most of the other crystallite orientations unpolarized in spite of the presence of the polarized electrons in their vicinity.

To polarize nuclei and harness the benefit of triplet-DNP in powder, it is necessary to circumvent the effect of serious anisotropic EPR broadening.
Kagawa et al. developed a UV-cured, magnetically oriented microcrystal array (MOMA) of pentacene-doped \textit{p}-terphenyl to eliminate the orientational distribution~\cite{kagawa2023triplet}.
To make the effect of anisotropic broadening small enough, Ajoy et al. reported optical hyperpolarization of $^{13}$C spins in diamond powder using NV$^{-}$ centers in such a low magnetic field of around 30~mT~\cite{Ajoy18PowderNVSweptFreqComm,Ajoy20RevSciInst} that the ZFS interaction is dominant over the Zeeman interaction with frequency-swept microwave irradiation over several gigahertz. 
The hyperpolarized $^{13}$C NMR signal was also observed in diamond powder with particle sizes of 2~$\mu$m and 100~nm at a magnetic field of 0.29~T for polarization and 1~T for NMR detection~\cite{BlinderSciAdv2025}. This was achieved by a combination of several techniques, including surface treatment, a microphotonic structure for laser irradiation, the PulsePol dynamic nuclear polarization sequence, and slow sample rotation.

In another approach, Chen et al. proposed DNP of $^{13}$C in nanodiamond powder through a double-quantum (DQ) transition, i.e., a transition associated with absorption/emission of two microwave photons between states with difference $|\Delta M_{S}|$ in the magnetic quantum number $M_{S}$ by $2$~\cite{chen2015}.
Since the DQ transition is free from anisotropy in first order, broadening is significantly scaled down.
One potential issue of triplet-DNP through the DQ transition is spectral overlap with the conventional single-quantum transition with $|\Delta M_S| = 1$.
For example, electron spins arising from another type of defect known as P1 centers undergo the single-quantum transition under the same condition as the DQ transition of the NV$^{-}$ centers.
Thus, DQ triplet-DNP can cause unintended side effects due to P1 centers which cannot be optically polarized.

\begin{figure}
  \includegraphics[width=0.45\textwidth]{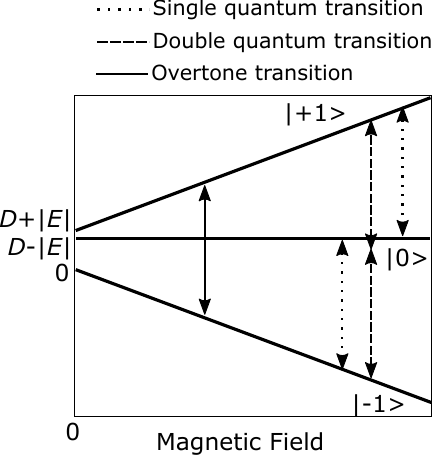}
  \caption{Magnetic-field dependence of the energies of the three states $|1 \rangle,|0 \rangle,| -1 \rangle$ of an electron spin in the triplet state. Vertical arrows with the same length indicate single quantum (dotted line), DQ (dashed line), and half-field or overtone (solid line) transition at a given, common microwave frequency.}
  \label{fig:pentacene_energy_diagram}
\end{figure}

Here, we report \textit{overtone} triplet-DNP, namely, DNP using optically polarized electron spins in the triplet state with the overtone transition, which refers to the $|\Delta M_S|= 2$ transition associated with absorption/emission of a \textit{single} microwave photon at a doubled frequency compared to the DQ transition in a given magnetic field ~\cite{shamesMagneticResonanceTracking2015,jeong2017,teradaMonodisperseFiveNanometerSizedDetonation2019,segawaHowIdentifyAttribute2020}.
Figure~\ref{fig:pentacene_energy_diagram} illustrates the distinction between DQ and overtone transitions.
We show Rabi oscillations between overtone states, which is, to our knowledge, the first experimental example of an overtone Rabi oscillation obtained from optically polarized triplet electron spins, proving the feasibility of coherent manipulation of spins between the overtone states that also makes coherent polarization transfer to nuclear spins possible.

Similarly to the DQ transition, the overtone transition is free from first-order broadening~\cite{tycko1987overtone,marinelli1999density}.
In the DQ transition, the otherwise forbidden overtone transition becomes allowed by the microwave field, whereas for the overtone transition, the perturbation introduced by the ZFS interaction plays a role.
For the overtone transition, the resonance condition is well separated from the single-quantum transition.

The overtone transition in NMR spectroscopy is well known as the \textit{half-field} transition in continuous-wave EPR, where the microwave excitation frequency is fixed while the magnetic field is swept to record a spectrum~\cite{van1959paramagnetic,de1960paramagnetic,weil2007electron}. In this setting, the $|\Delta M_S|= 2$ transition shows a resonant signal at about \textit{half} the magnetic field, compared to the fully allowed $|\Delta M_S|= 1$ single quantum transitions (see Fig.~\ref{fig:pentacene_energy_diagram}).
In contrast, for pulse NMR spectroscopy of spin-1 nuclei, the term overtone spectroscopy has become established since when using a fixed magnetic field, the nuclear spins must be excited at twice the fundamental single-quantum frequency to match the $|\Delta M_S|= 2$ (i.e., first) \textit{overtone} transition~\cite{tycko1987overtone,marinelli1999density}.
We adopt the term \textit{overtone}, since we rely on the theoretical description in the literature on overtone NMR spectroscopy~\cite{tycko1987overtone,marinelli1999density}.

In the following, we describe a theoretical framework of the overtone transition and show experimental observation of the overtone Rabi oscillation of the electron spins in the photo-excited triplet state of pentacene as well as in the optically polarized ground triplet state of NV$^-$ centers in microdiamonds.
Then, proton hyperpolarization is demonstrated in polycrystalline pentacene-doped \textit{p}-terphenyl using the integrated solid effect (ISE)\cite{henstra1990high} as a polarization mechanism applied to the overtone transition.

\section{Theory}\label{sec:theory}
\subsection{Resonance condition and nutation rate of overtone transition}
Let us consider an electron spin $S=1$ in the triplet state in a static magnetic field $\bm{B}_{0} \parallel z$ and subject to a microwave field $\bm{B}_{1}$ at frequency $\omega_{\mathrm{MW}}$ close to the overtone transition frequency $2\omega_{e}$, where $\omega_{e}$ is the Larmor precession frequency given with the gyromagnetic ratio $\gamma_{e}$ as
\begin{align}
  \omega_{\mathrm{e}} = - \gamma_{e} B_{0}.
  \label{eq:larmorfreq}
\end{align}
We write the Hamiltonian $H$ of the system as
\begin{align}
    H = H_{\mathrm{Z}} + H_{\mathrm{MW}} + H_{\mathrm{ZFS}}.
\end{align}
Here, $H_{\mathrm{Z}}$ and $H_{\mathrm{MW}}$ are the Zeeman interaction with the static and the microwave fields
\begin{align}
  H_{\mathrm{Z}} &= \omega_{\mathrm{e}} S_z, \\
  H_{\mathrm{MW}} &= 2\omega_{1} \left[ \sin \chi S_x+\cos \chi S_z \right] \cos (\omega_{\mathrm{MW}} t ),
\end{align}
where $\chi$ is the angle between the static field $\bm{B}_{0}$ and the microwave field $\bm{B}_{1}$, and
\begin{align}
    \omega_{1} = - \frac{1}{2} \gamma_{e} B_{1}
\end{align}
is the single-quantum nutation frequency.
The ZFS interaction $H_{\mathrm{ZFS}}$ is expressed as
\begin{align}
    H_{\mathrm{ZFS}} = \mathrm{\omega_{ZFS}} \left[3 S_{z}^{2} - S(S+1)  + \eta \left( S_{x}^{2} - S_{y}^{2} \right)\right],
    \label{eq:hzfs}
\end{align}
where $\omega_{\mathrm{ZFS}}$ and $\eta$ are related to the ZFS parameters $D$ and $E$ by
\begin{align}
    \omega_{\mathrm{ZFS}} &= \frac{D}{3}, \\
    \eta &= \frac{E}{\omega_{\mathrm{ZFS}}}.
\end{align}

In the case of electron spin in the photo-excited triplet state of pentacene doped with $p$-terphenyl, $D$ and $E$ are $2\pi\cdot 1395.57$~MHz and $2\pi\cdot 53.35$~MHz~\cite{yang2000zero}, while for the NV$^-$ center in diamond, $D$ and $E$ are $2\pi\cdot 2870$~MHz and $\approx0$~\cite{doherty2013nitrogen}.
In a magnetic field of the order of 0.2~T used in this work for the overtone experiments, the electron Larmor frequency $\omega_e=2\pi\cdot 5.8$~GHz is much higher than $\omega_{\mathrm{ZFS}}\simeq 2\pi\cdot 465$~MHz for pentacene and $\omega_{\mathrm{ZFS}}\simeq 2\pi\cdot 957$~MHz for NV$^-$ centers.
To appreciate that $\omega_{\mathrm{ZFS}}$ is considerably lower than $\omega_{e}$, we introduce
\begin{align}
    \varepsilon \equiv \frac{\omega_{\mathrm{ZFS}}}{\omega_{e}} \ll 1,
\end{align}
which is ca.~0.08 and 0.165 for pentacene and NV$^{-}$ centers.

Following a formalism of overtone NMR for $^{14}$N (spin 1) ~\cite{tycko1987overtone,marinelli1999density} where the quadrupolar interaction comes into play and has the same form as Eq.~(\ref{eq:hzfs}), we introduce dimensionless functions, $f$, $g$, and $h$, of the Euler angles $\{ \alpha, \beta, \gamma\}$ that connect the principal axis system of the ZFS tensor (instead of the quadrupolar tensor) and the laboratory coordinates as
\begin{align}
    f(\alpha,\beta,\gamma) =& e^{i\gamma}[3\sin\beta \cos\beta \nonumber \\
                & -\eta(\cos 2\alpha \sin\beta \cos\beta + i \sin 2\alpha \sin \beta) ],  \\
    g(\alpha,\beta,\gamma) =& e^{2i\gamma}\frac{1}{2}[3\sin^2\beta 
                 +\eta(\cos 2\alpha (1+\cos^2\beta) \nonumber \\ 
                 & + 2i\sin2\alpha\cos\beta) ], \\ 
    h(\alpha,\beta,\gamma) =& f f^{*} + g g^{*}.
\end{align}
Then, the resonance condition of the overtone transition is obtained as
\begin{align}
    \omega_{\mathrm{MW}} = 2 \omega_{e} + 2 \varepsilon \omega_{\mathrm{ZFS}} h,
    \label{eq:resonance-condition}
\end{align}
and the on-resonance overtone nutation frequency $\omega_{\mathrm{nut}}$ as
\begin{align}
    \omega_{\mathrm{nut}} = \varepsilon \omega_{1} | f \sin\chi + g \cos\chi|.
    \label{eq:nutation-freq}
\end{align}
The derivation of Eqs.~(\ref{eq:resonance-condition})-(\ref{eq:nutation-freq}) is described in the Supplemental Material~\cite{supp} (see also references \cite{luttinger1955motion,schrieffer1966relation} therein).

\subsection{\label{subsec:lineshape}Anisotropic shift of overtone resonance}
The resonance frequency of the overtone transition is not exactly twice the electron Larmor frequency, but has a shift according to the second term of Eq.~(\ref{eq:resonance-condition}).
Since the shift depends on the orientation of the individual crystallites through $h(\alpha,\beta,\gamma)$, the random distribution of the Euler angles in the powder causes the resonance line to broaden.
Importantly, the broadening scales as $\varepsilon\omega_{\mathrm{ZFS}}=\omega_{\mathrm{ZFS}}^{2}/\omega_{e}$, and is therefore expected to be much smaller than the anisotropic broadening of the single-quantum transition.

For NV$^{-}$ centers in diamond, the ZFS tensor is axially symmetric ($\eta=0$), and for electrons in the photo-excited triplet state of pentacene, $\eta \ll 1$.
In this work, we derived a lineshape function $I(\omega)$ of the overtone spectrum in the limit of $\eta \rightarrow 0$ to be (see Supplemental Material~\cite{supp})
\begin{widetext}
\begin{align}
  I(\omega) = 
       \frac{1}{18} \sqrt{\frac{3}{2}} \frac{1}{\varepsilon\omega_{\mathrm{ZFS}}} \left[ 
       \frac{6\varepsilon\omega_{\mathrm{ZFS}}}{2\omega_{e}+6\varepsilon\omega_{\mathrm{ZFS}}-\omega} \right]^{\frac{1}{2}}
       \left[ 1 + 2 \sqrt{\frac{2\omega_{e}+6\varepsilon\omega_{\mathrm{ZFS}}-\omega}{6\varepsilon\omega_{\mathrm{ZFS}}} } \right]^{-\frac{1}{2}},
       \label{eq:iw1}
\end{align}
for $2\omega_{e} \le \omega < 2\omega_{e}+\frac{9}{2}\varepsilon\omega_{\mathrm{ZFS}}$,
\begin{align}
  I(\omega) =& 
       \frac{1}{18} \sqrt{\frac{3}{2}} \frac{1}{\varepsilon\omega_{\mathrm{ZFS}}} \left[ \frac{6\varepsilon\omega_{\mathrm{ZFS}}}{2\omega_{e}+6\varepsilon\omega_{\mathrm{ZFS}}-\omega} \right]^{\frac{1}{2}} \nonumber \\
       & \times
       \left\{ 
       \left[ 1 + 2 \sqrt{\frac{2\omega_{e}+6\varepsilon\omega_{\mathrm{ZFS}}-\omega}{6\varepsilon\omega_{\mathrm{ZFS}}} } \right]^{-\frac{1}{2}}
       +
       \left[ 1 - 2 \sqrt{\frac{2\omega_{e}+6\varepsilon\omega_{\mathrm{ZFS}}-\omega}{6\varepsilon\omega_{\mathrm{ZFS}}} } \right]^{-\frac{1}{2}}
       \right\},
       \label{eq:iw2}
\end{align}
\end{widetext}
for
$2\omega_{e}+\frac{9}{2}\varepsilon\omega_{\mathrm{ZFS}} < \omega < 2\omega_{e} + 6 \varepsilon\omega_{\mathrm{ZFS}}$,
and $I(\omega) = 0$ for other regions of $\omega$.

Inspecting Eqs.~(\ref{eq:iw1})-(\ref{eq:iw2}) and Fig.~S1, we find singularities in $I(\omega)$ at $\omega= 2\omega_{e}+\frac{9}{2}\varepsilon\omega_{\mathrm{ZFS}}$ and $\omega= 2\omega_{e}+6\varepsilon\omega_{\mathrm{ZFS}}$.
This would be the case if the individual packets of the resonance line were given by a delta function, whereas in reality, they cannot be infinitely sharp, so that $I(\omega)$ should look like a convolution with a finite width function.
The figure also indicates that, even though the resonance condition is distributed from $2\omega_{e}$ to $2\omega_{e}+6\varepsilon\omega_{\mathrm{ZFS}}=2\omega_{e}+2 \varepsilon D$, most resonances concentrate in the narrower region between the two singularities ranging over $\frac{3}{2}\varepsilon\omega_{\mathrm{ZFS}}=\frac{1}{2}\varepsilon D$ centered at $2\omega_{e} + \frac{21}{4}\varepsilon\omega_{\mathrm{ZFS}}=2\omega_{e}+\frac{7}{4}\varepsilon D$.

In the case of field-variable experiments with a \textit{fixed} frequency $\omega_{\mathrm{MW}}$, the profile of the lineshape $I(B_{0})$ as a function of the applied static field $B_{0}$ is obtained by inserting Eq.~(\ref{eq:larmorfreq}) into Eqs.~(\ref{eq:iw1})-(\ref{eq:iw2}), keeping in mind that $\omega=\omega_{\mathrm{MW}}$ is now a constant and that $\varepsilon$ implicitly includes $B_{0}$.
In the field-variable case, we obtain the shift $B_{\mathrm{s}}$ of resonance from the exact overtone condition $\omega_{\mathrm{MW}}/(2|\gamma_{e}|)$ to be
\begin{align}
    B_{\mathrm{s}} = 
      -\frac{\omega_{\mathrm{MW}}}{|\gamma_{e}|} \frac{21}{16} \varepsilon^{2},
\end{align}
and the width $B_{\mathrm{w}}$ be
\begin{align}
    B_{\mathrm{w}} = 
      \frac{\omega_{\mathrm{MW}}}{|\gamma_{e}|} \frac{3}{8} \varepsilon^{2}.
\end{align}
Thus, we expect that the width amounts to $\frac{6}{21} \approx 0.29$ times the shift.
With $\varepsilon \approx 0.08~(0.165)$ for pentacene~(NV$^{-}$) and $\omega_{\mathrm{MW}} = 2\pi \cdot 11.6$~GHz used in this work, we anticipate that the shift $B_{\mathrm{s}}$ to be ca.~-3.5 mT and ca.~-15 mT for pentacene and NV, respectively.

\subsection{\label{subsec:nutation-anisotropy}Anisotropic overtone nutation}
Eq.~(\ref{eq:nutation-freq}) tells that the overtone nutation frequency $\omega_{\mathrm{nut}}$ is also anisotropic and scales as $\varepsilon \omega_{1}$.
In addition, the overtone transition is induced by both perpendicular ($\chi=\pi/2$) and parallel ($\chi=0$) resonant microwave fields, which is also well known in $^{14}$N overtone NMR~\cite{tycko1987overtone,marinelli1999density}. 
In the conventional case of $\bm{B}_{0} \perp \bm{B}_{1}$, the nutation frequency $\omega_{\mathrm{nut},\perp}$ is, in the limit of $\eta \rightarrow 0$,
\begin{align}
    \omega_{\mathrm{nut},\perp} = 3\varepsilon\omega_{1}\sin\beta\cos\beta.
    \label{eq:omeganut-perp}
\end{align}
Thus, the nutation rate takes the maximum value of $\frac{3}{2}\varepsilon\omega_{1}$ at $\beta=\frac{\pi}{4}$.
In contrast, $\beta=0$ and $\frac{\pi}{2}$ lead to zero nutation frequency, so the ZFS tensor that happens to have these orientations is expected to have little contribution to EPR and triplet-DNP.

As we demonstrate below, the experimental nutation frequencies in our studies were below $2\pi\cdot 1$~MHz, much lower even than the \textit{residual} broadening $(\approx \varepsilon D)$ of the overtone resonance. If the microwave pulse were made strong enough in the future to excite all packets of overtone resonance, the overall effect of anisotropic nutation would be described by the distribution function $I_{\perp}(\omega_{\mathrm{nut}})$
\begin{widetext}
\begin{align}
    I_{\perp}(\omega_{\mathrm{nut}}) =
    \frac{1}{3\sqrt{2}\varepsilon\omega_{1}}
    \left[ 1 - 4 \left( \frac{\omega_{\mathrm{nut}}}{3\varepsilon\omega_{1}} \right)^{2}\right]^{-\frac{1}{2}} 
    \left[ 
    \sqrt{1-\sqrt{1-4\left(\frac{\omega_{\mathrm{nut}}}{3\varepsilon\omega_{1}} \right)^{2}}} 
    +
    \sqrt{1+\sqrt{1-4\left(\frac{\omega_{\mathrm{nut}}}{3\varepsilon\omega_{1}} \right)^{2}}} 
    \right].
    \label{eq:nut-perp}
\end{align}    
\end{widetext}
Eq.~(\ref{eq:nut-perp}) is derived in the Supplemental Material~\cite{supp}.
It is found from Eq.~(\ref{eq:nut-perp}) that $I_{\perp}(\omega_{\mathrm{nut}})$ diverges at $\omega_{\mathrm{nut}}=\frac{3}{2}\varepsilon\omega_{1}$.
The nutation frequency $\omega_{\mathrm{nut},\parallel}$ in the case of $\bm{B}_{0} \parallel \bm{B}_{1}$ is, again in the $\eta \rightarrow 0$ limit,
\begin{align}
    \omega_{\mathrm{nut},\parallel} = \frac{3}{2} \varepsilon \omega_{1}\sin^{2}\beta,
\end{align}
whence we obtain the distribution function $I_{\parallel}(\omega_{\mathrm{nut}})$ (see Supplemental Material~\cite{supp}) to be
\begin{align}
    I_{\parallel}(\omega_{\mathrm{nut}}) = \frac{1}{3\varepsilon\omega_{1}}
    \left[1 - \frac{2\omega_{\mathrm{nut}}}{3\varepsilon\omega_{1}} \right]^{-\frac{1}{2}}.
\end{align}

In both of the above cases, the overtone nutation does not occur for $\beta=0$.
However, $\beta=0$ is the least populated orientation of the Euler angle $\beta$ due to the weight factor given by $\sin\beta$, so that the zero nutation frequency arising from $\beta=0$ would have little effect on the EPR and triplet-DNP experiments.
For the most populated orientation, $\beta=\pi/2$, $\omega_{\mathrm{nut},\perp}$ vanishes, but $\omega_{\mathrm{nut},\parallel}$ does not.
Thus, experiments with $\bm{B}_{1} \parallel \bm{B}_{0}$ can complement inefficient nutation with $\bm{B}_{1} \perp \bm{B}_{0}$.
The price to pay would be that a special, non-conventional arrangement of the microwave resonator is required which may not be straightforwardly feasible in the conventional equipment.

\section{Experimental}
0.05~mol\% pentacene-doped $p$-terphenyl was prepared from $p$-terphenyl purchased from Tokyo Chemical Industry and pentacene from Sigma-Aldrich.
We henceforth refer to the sample as PHPT, with natural-abundance pentacene (P) molecules with hydrogen (H) atoms doped in $p$-terphenyl (PT), in contrast with deuterated-pentacene-doped $p$-terphenyl, dubbed as PDPT. The structural formulas of these molecules are shown in Fig.~S3.
The microdiamond sample is the same as the one we used previously~\cite{miyanishi2021room}, with 8.9$\times$10$^{17}$ cm$^{-3}$ (5~ppm in atomic ratio) NV$^-$ concentration and $4.6\pm0.1\times10^{18}$ cm$^{-3}$ (26~ppm) P1 center concentration. The size of the microdiamond is 500~$\mu$m.
All samples were packed in a glass tube.

The experimental setup used in this work is similar to that reported in~\cite{miyanishi2021room}, and is described in the Supplemental Material~\cite{supp} in detail.
A cylindrical cavity was placed in an electromagnet with a static magnetic field ranging from 0.19 to 0.5~T. To irradiate microwaves perpendicular or parallel to the static magnetic field, the cylindrical cavity was fixed in two different orientations relative to the static magnetic field (shown in Supplemental Material~\cite{supp} Fig.~S2).
The static magnetic field was generated by an electromagnet and applied to the samples placed in a home-built X-band cavity resonating at 11.6~GHz.
A solid-state laser with a wavelength of 527~nm, pulse length of 200~ns, and pulse energy of 30~mJ was employed for a light source for photo-excitation of NV$^{-}$ centers, and a dye laser with a wavelength of 594~nm, pulse length of 200~ns, and pulse energy of 6~mJ was used for photo-excitation of pentacene.
EPR measurements with a spin-echo sequence (Fig.~\ref{fig:EPR_measurments_pentacene}(a)) were performed on a home-built spectrometer equipped with a superheterodyne architecture capable of converting microwave pulses at 2.304~GHz, generated by a Xilinx ZCU111 RFSoC Evaluation Board with software referred in~\cite{rftool_client}, into those at 11.6~GHz with a power of 0.44~W.
In DNP experiments, microwave pulses at 11.6~GHz, generated by Vaunix Lab Brick LMS 183DX, were amplified to ca.~19~W using a solid state high power amplifier~(AMP2036, Exodus Advanced Communications).
All experiments were carried out at room temperature.

\section{Results and discussion}

\subsection{EPR measurements}

\begin{figure}[h]
  \includegraphics[width=0.4\textwidth]{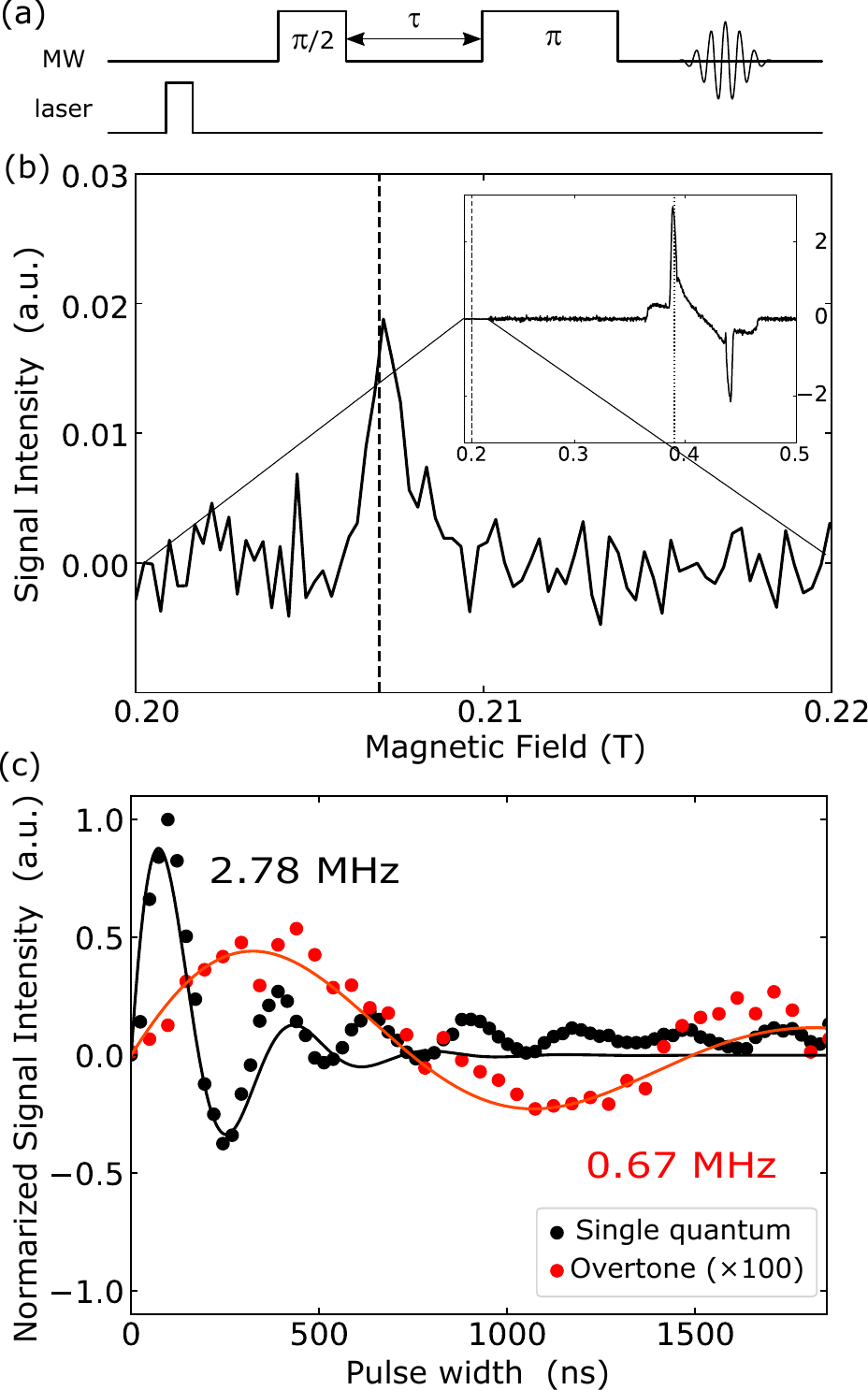}
  \caption{(a) EPR echo pulse sequence used in the measurement. $\tau$ is the echo time and set to 1~$\mu$s. For the Rabi measurement, the pulse length of $\pi/2$ is swept, and the microwave frequency is 11.6~GHz. (b) Powder EPR spectra of pentacene doped $p$-terphenyl. The spectrum from 0.22~T to 0.5~T (inset) was obtained with accumulation over 100 times, and the pulse lengths for $\pi/2$ and $\pi$ were adjusted to maximize the single-quantum EPR signal. The spectrum ranging from 0.2~T to 0.22~T was obtained with 5,000 scans with the pulse lengths set to maximize the overtone EPR signal. (c) Experimental Rabi oscillations (spin echo amplitude as a function of the width of the first microwave pulse) measured at 0.39~T (dotted line in the inset figure in (b)) corresponding to the single quantum transition (black circles) and at 0.207~T (dashed line in (b)) for the overtone transition (red circles). The oscillation signal for single quantum transition was obtained with accumulation over 100 times, whereas that for overtone transition was obtained with accumulation over 5,000 times. Solid lines are fitted curves of a decaying sinusoidal function ($a\exp(-\Gamma t)\sin(\omega_{\mathrm{nut}} t)$). The signal intensity obtained for the overtone transition is scaled by a factor of 100. The $\pi$ pulse length in these measurements were 175~ns for the single quantum transition and 585~ns for the overtone transition.}
  \label{fig:EPR_measurments_pentacene}
\end{figure}

Figure~\ref{fig:EPR_measurments_pentacene}(b) shows the magnetic-field dependence of the amplitudes of the spin echoes of the electron spins in the photo-excited triplet state of pentacene in a 10~mg PHPT sample.
The single quantum transition was found in magnetic fields ranging from ca.~0.36 to 0.47~T over ca.~0.11~T, showing anisotropy of both the resonance condition and the spin polarization~\cite{miyanishi2021room}.
In addition, we found a much weaker resonance at 0.207~T, which can be assigned to the overtone transition.
The full width at half maximum~(FWHM) was ca.~1~mT, which was narrower by two orders of magnitude compared to the single quantum transition.
The spectrum from 0.22~T to 0.5~T was obtained with accumulation over 100 times, and that from 0.2~T to 0.22~T was obtained with 5,000 scans.

Fig.~\ref{fig:EPR_measurments_pentacene}(c) demonstrates single-quantum and overtone Rabi nutations obtained with incremented width of the first microwave pulse in Fig.~\ref{fig:EPR_measurments_pentacene}(a).
The solid lines represent the fitting of the experimental data with an exponentially decaying sinusoidal function $a\sin(\omega_{\mathrm{nut}}t) e^{-\Gamma t}$, where $a$, $\Gamma$, and $\omega_{\mathrm{nut}}$ \textit{roughly} measure the amplitude, the decay rate, and the nutation frequency. Some discrepancy, seen in particular for the single-quantum case (black line), suggests that the choice of the fitting function was less than ideal. Indeed, in both the single-quantum and the overtone experiments, we expect that multiple Rabi frequencies are involved in such a powder sample. In addition, spin decoherence and lifetime decay of the individual sub-levels of the triplet state also affect the profile of the Rabi oscillation in a rather complicated way. Obviously, an accurate reproduction of the experimental data would require an elaborate model that takes these factors into account. Leaving the challenge to our future studies, we are content here with a crude but convenient order estimation of the intensity $\omega_{\mathrm{nut}}$ of the microwave field felt by the electron spin in the triplet state from the curve fitting.
The Rabi nutation frequencies of the single quantum transition and the overtone transition were ca.~$2\pi\cdot 2.78$~MHz and $2\pi\cdot 0.67$~MHz with the same applied microwave power.
The lower nutation rates and the smaller signal amplitude for the overtone transition were in line with the theory described in Section~\ref{sec:theory} suggesting they both scale with $\varepsilon$.

\begin{figure}
  \includegraphics[width=0.4\textwidth]{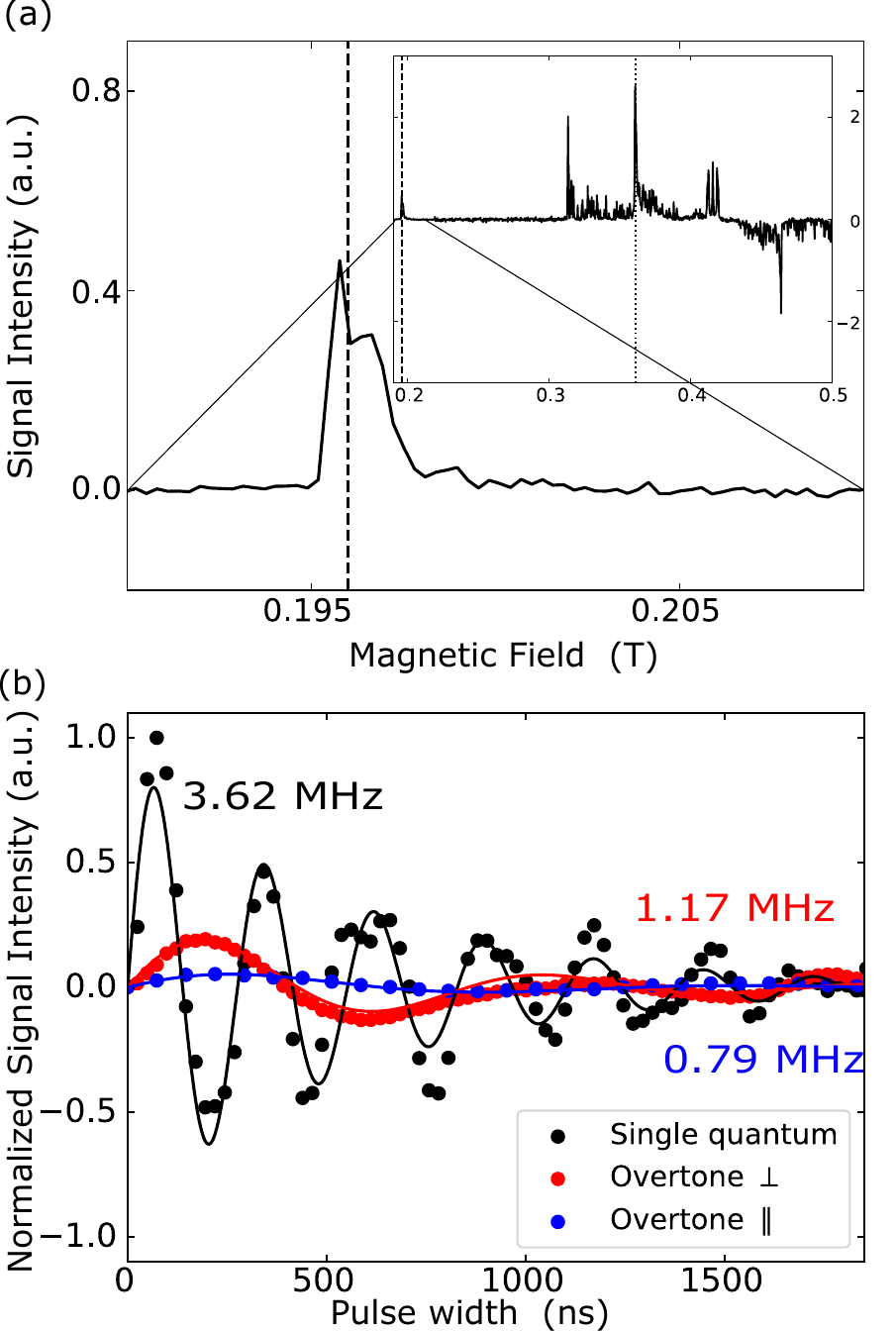}
  \caption{(a) An EPR spectrum of NV$^-$ centers in microdiamond powder measured at 11.6~GHz. In magnetic fields ranging from 0.22~T to 0.5~T data were obtained with the lengths of the $\pi/2$ and the $\pi$ pulses set to maximize the single-quantum EPR signal with 100 times signal averaging (inset). The spectrum ranging from 0.19~T to 0.22~T was obtained with 1,000 times signal averaging and the pulse lengths adjusted to maximize the overtone EPR signal. (b) Experimental Rabi oscillations for single-quantum transition (black circles) at 0.361~T (dotted line in the inset figure in (b)) and overtone transition with $\bm{B}_{1} \perp(\parallel) \bm{B}_{0}$ (red(blue) circles) at 0.196~T (dashed line in (b)). The oscillation signal for single quantum transition at 0.361~T was obtained with accumulation over 100 times, whereas that for overtone transitions at 0.196~T was obtained with accumulation over 1,000 times. The solid lines are fitted curves with a decaying sinusoidal function. The $\pi$ pulse length in these measurement were 146~ns for the single quantum transition, 391~ns for the overtone transition with $\bm{B}_{1} \perp \bm{B}_{0}$, and 488~ns for overtone transition with $\bm{B}_{1} \parallel \bm{B}_{0}$.}
  \label{fig:EPR_measurments_diamond}
\end{figure}

Similar experiments were performed on NV$^-$ centers in microdiamonds weighing 40~mg.
Figure~\ref{fig:EPR_measurments_diamond}(a) shows the magnetic field dependence of the spin echoes of the electrons in the triplet state measured after photo-excitation.
We observed spin-echo signals from the single quantum transition in a wide range of magnetic fields from 0.22~T to 0.5~T due to the anisotropy of the ZFS interaction.
The number ($ \approx 200$) of the microdiamond crystallites was too small to show a smooth powder spectrum.
The three lines in the center of the field at 0.416~T stem from P1 centers and other $S$ = 1/2 electron spin defects within the microdiamonds~\cite{miyanishi2021room}.
In addition, we found a much narrower resonance of the overtone transition, shifted from the half-field condition (0.207~T) by ca.~-11~mT with a width of ca.~1.7~mT, both of which are consistent with the theoretical estimation that we discussed in section~II.
The spectrum from 0.22~T to 0.5~T was obtained with accumulation over 100 times, and that from 0.19~T to 0.22~T was obtained with 1,000 scans.

As shown in Fig.~\ref{fig:EPR_measurments_diamond}(b), the single-quantum transition at 0.361~T underwent nutation at a rate of $2\pi\cdot 3.62$~MHz, while the overtone nutation frequency $\omega_{\mathrm{nut},\perp}$ at 0.196~T with the microwave field $\bm{B}_{1}$ set perpendicular to the static field $\bm{B}_{0}$ was $2\pi\cdot 1.17$~MHz, about three times smaller than the former for the same microwave amplitude. 
According to Eq.~(\ref{eq:omeganut-perp}), the overtone nutation rate $\omega_{\mathrm{nut},\perp}$ is lower than the single-quantum nutation rate by a factor $3\varepsilon\sin\beta\cos\beta$.
With $\varepsilon=0.165$ for NV$^{-}$ and distribution of the angle $\beta$, the experimental scaling factor (0.32) was reasonable.
The overtone Rabi frequency for NV$^-$ nearly twice that of pentacene can be explained by the ratio of the $\varepsilon$ values of $(0.165/0.08)$. 
We also successfully measured the overtone Rabi nutation with an experimental setup in which the resonant microwave field $\bm{B}_{1}$  is parallel to the static magnetic field $\bm{B}_{1}$ (Fig.~\ref{fig:EPR_measurments_diamond}(b)). The overtone nutation frequency $\omega_{\mathrm{nut},\parallel}$ was $2\pi\cdot 0.79$~MHz.

\begin{figure}
  \includegraphics[width=0.45\textwidth]{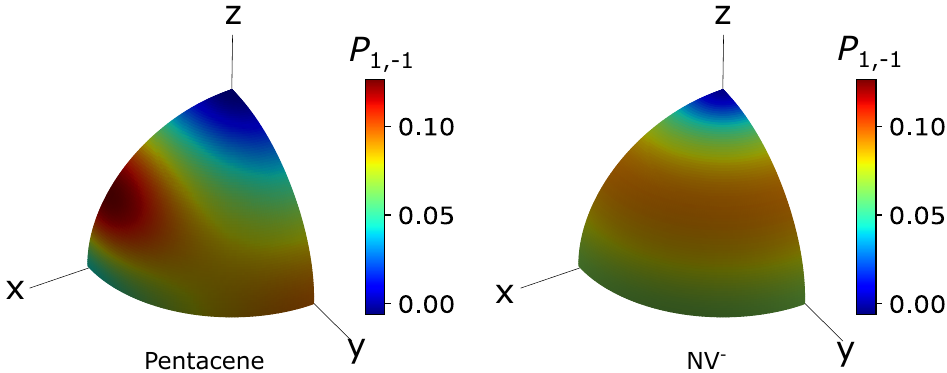}
  \caption{Orientational dependence of polarization $P_{1,-1}$ between the overtone states for pentacene doped in $p$-terphenyl ($B=0.207$~T) and NV$^-$ center ($B=0.196$~T).}
  \label{fig:polarization_orientation}
\end{figure}

\begin{table*}[htb]
  \begin{tabular}{ccc} \hline \hline
    Transition & Single quantum & Overtone ($\bm{B}_{0} \perp \bm{B}_{1}$) \\ \hline
    density of pentacene & \multicolumn{2}{c}{$2.5\times10^{18}$cm$^{-3}$~(0.04~mol$\%$)}   \\ \hline
    ZFS parameter $D$~/~MHz & \multicolumn{2}{c}{1395.57} \\ \hline
    ZFS parameter $E$~/~MHz & \multicolumn{2}{c}{53.35}  \\ \hline
    $B_{0}$~/~T & 0.36-0.47 & 0.207 \\ \hline
    Rabi nutation frequency~/~MHz & 2.78 & 0.67 \\ \hline
    \hline
  \end{tabular}
    \caption{Summary of sample condition and the parameters for the EPR experiments using pentacene.}
\label{Table:ParametersForSampleAndOptimisedSequence}
\end{table*}

\begin{table*}[htb]
  \begin{tabular}{cccc} \hline \hline
    Transition & Single quantum & \quad Overtone ($\bm{B}_{0} \perp \bm{B}_{1}$) & \quad Overtone ($\bm{B}_{0} \parallel \bm{B}_{1}$) \\ \hline
    density of NV$^{-}$ & \multicolumn{3}{c}{$8.9\times 10^{17}$~cm$^{-3}$~(5~ppm)} \\ \hline 
    ZFS parameter $D$~/~MHz & \multicolumn{3}{c}{2870}   \\ \hline
    ZFS parameter $E$~/~MHz & \multicolumn{3}{c}{0}   \\ \hline
    $B_{0}$~/~T & 0.31-0.50 & \multicolumn{2}{c}{0.196} \\ \hline
    Rabi nutation frequency~/~MHz & 3.62 & 1.17 & 0.79 \\ \hline
    \hline
  \end{tabular}
    \caption{Summary of sample condition and the parameters for the EPR experiments using NV$^-$.}
    \label{Table:ParametersForEPR_NV}
\end{table*}

Regarding the intensity of the overtone EPR signals, the NV$^-$ sample showed a 25.5 times larger echo compared to pentacene.
The signal intensity depends on factors such as the amount of sample undergoing photoexcitation, the electron polarization between the overtone energy levels, and the scaling factor $\varepsilon$. 
We compared EPR signal intensities with such a tiny sample that all pentacene molecules are guaranteed to be photo-excited with our laser, and estimated that about 21.3\% of the pentacene molecule were photo-excited while all NV$^-$ centers were.
Taking also the concentration and amount of sample into account, we found the ratio of the optically polarized electron spins in microdiamonds and to those in PHPT powder to be 7.2.
Calculated orientational dependence of the overtone polarization $P_{1,-1}$ for pentacene and NV$^{-}$ are shown in Fig.~\ref{fig:polarization_orientation}.
For pentacene, the population fractions $\rho_{X}$, $\rho_{Y}$, and $\rho_{Z}$ over the ZFS eigenstates $|X\rangle$, $|Y\rangle$, and $|Z\rangle$ were set to $\rho_{X}:\rho_{Y}:\rho_{Z}=0.76:0.16:0.08$~\cite{sloop1981electron,ong1995deuteration} in the calculation. For NV$^{-}$ centers, the populations $\rho_{0}$ and $\rho_{\pm 1}$ over the symmetric ZFS eigenstates with $m_{s}=0$ and $m_{s}=\pm 1$ are determined by the balance between non-selective optical pumping and selective intersystem crossing. Upon photo-excitation by \textit{our} pulsed laser, the single-quantum polarization of 0.16 was experimentally obtained at 0.361~T (for which the major principal axis of the ZFS tensor is normal to the field), whence we estimated $\rho_{0}=0.46\pm0.02$ and $\rho_{\pm 1}=0.54\pm0.02$ (see Supplemental Material~\cite{supp} and also reference~\cite{miyanishi2021room} therein).
From the results in Fig.~\ref{fig:polarization_orientation}, the average polarizations $P_{1,-1}$ of the overtone transition for NV$^-$ and pentacene were found to be 0.074 and 0.071.
Taking into account these differences in the hyper overtone polarization $P_{1,-1}$, the fraction $\rho_{1}+\rho_{-1}$, and the scaling factor $\varepsilon$, we estimated that the intensity of the EPR signal of NV$^-$ to be ca.~20 times that of pentacene, which is consistent with the value (25.5) obtained in the experiment.

The population in the central state is irrelevant to the overtone transitions. Nevertheless, it could be exploited in future to enhance the overtone polarization by applying population inversion between the central state and one of the other states.

\subsection{Hyperpolarized $^1$H NMR}
\begin{figure*}[t]
  \includegraphics[width=0.7\textwidth]{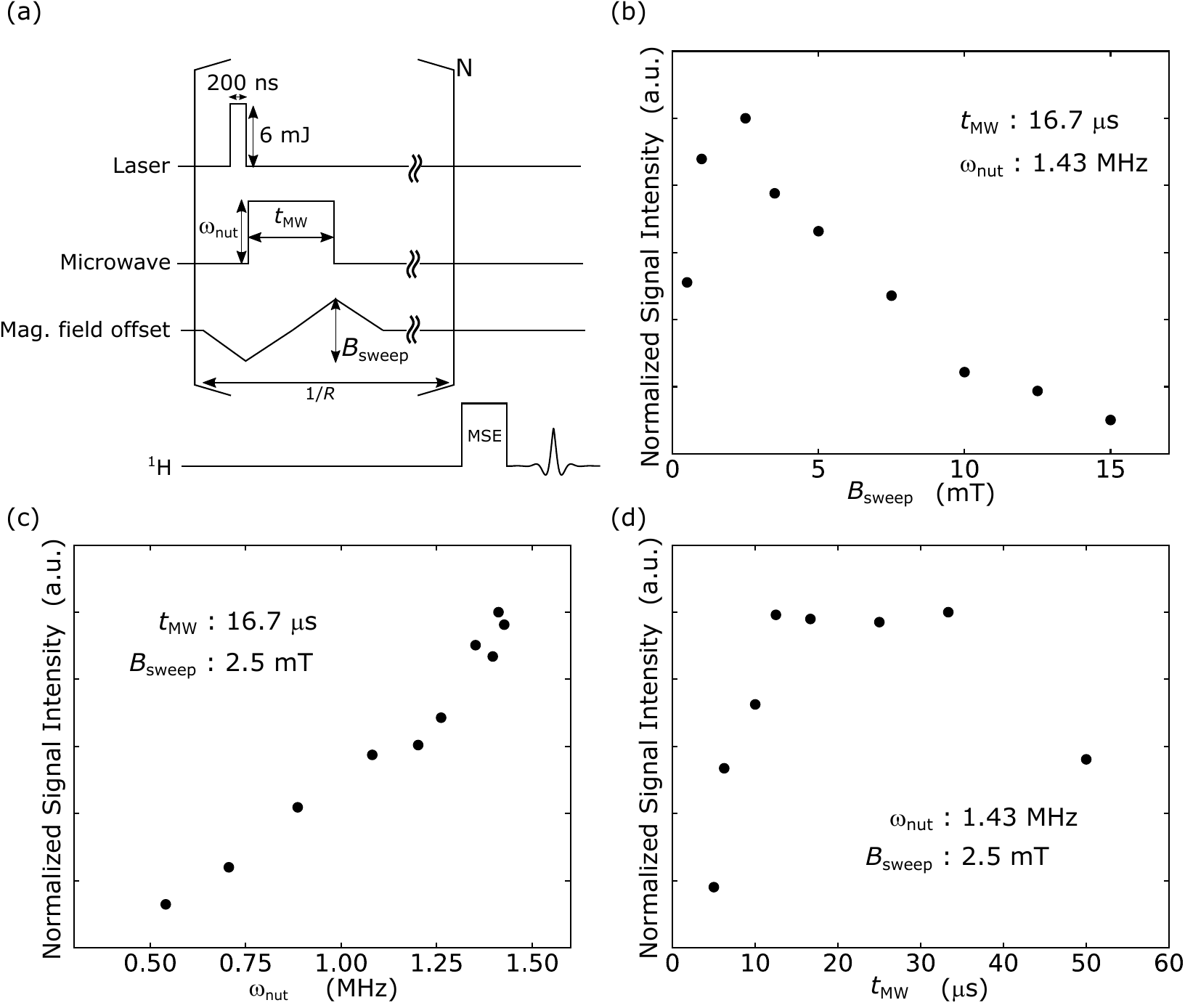}
  \caption{(a) Pulse sequence for the ISE followed by $^1$H NMR detection using MSE sequence. A laser pulse is used for photoexcitation of the triplet electron spins. Then, a microwave pulse is applied together with a magnetic-field sweep. The ISE pulse sequence is repeated before the enhanced $^1$H magnetization is detected by a MSE sequence. 
  The definitions of the symbols used in the figure are as follows. $\omega_{\mathrm{nut}}$: microwave Rabi nutation frequency, $t_{\mathrm{MW}}$: microwave-pulse width, $B_{\mathrm{sweep}}$: field-sweep width, $R$: ISE repetition rate, and N: ISE repetition number.
  (b-d) Optimization of the ISE pulse sequence for pentacene-doped $p$-terphenyl (PHPT) through adjustment of (b) the field-sweep width $B_{\mathrm{sweep}}$ sweep, (c) the microwave intensity $\omega_{\mathrm{nut}}$, and (d) the microwave duration $t_{\mathrm{MW}}$. 
  The values of the fixed parameters are indicated in each graph.
}
  \label{fig:DNP_optimisation}
\end{figure*}

In the magnetic field of 0.207~T at which overtone EPR was successfully observed, the ISE sequence was performed to hyperpolarize the $^1$H spins in a PHPT sample with a weight of 10~mg.
To find the experimental parameters that maximize the efficiency of the overtone triplet-DNP, we examined the enhanced $^1$H magnetization with a varied width $B_{\mathrm{sweep}}$ of magnetic-field sweep, the amplitude $\omega_{\mathrm{1}}$ of the microwave pulse, and the duration $t_{\mathrm{MW}}$ of the ISE sequence.
As demonstrated in Fig.~\ref{fig:DNP_optimisation}(b–d), the optimal conditions were found to be $B_{\mathrm{sweep}}$=2.5~mT, $t_{\mathrm{MW}}$=16.7~$\mu$s.
The overtone triplet-DNP efficiency increased with microwave amplitude $\omega_{\mathrm{1}}$ and did not saturate with the maximum available power of the microwave pulse of ca.~19~W (Fig.~\ref{fig:DNP_optimisation}(c)).
This indicates that the overtone Rabi frequency $\omega_{\mathrm{nut}} \sim \varepsilon \omega_{1}$ was considerably below the proton Larmor precession frequency $\omega_{\mathrm{0,n}}$ of $2\pi\cdot 8.74$~MHz, and the Hartmann-Hahn condition
\begin{align}
\label{eq:HH}
\omega_\mathrm{nut}=\omega_{\mathrm{0,n}}
\end{align}
was fulfilled with a significant resonance offset $\Delta \omega$ provided by the magnetic-field sweep as well as by the hyperfine coupling.
A more powerful microwave amplifier such as a traveling-wave tube (TWT) amplifier could improve the efficiency of overtone triplet-DNP.

\begin{figure*}
\includegraphics[width=0.7\textwidth]{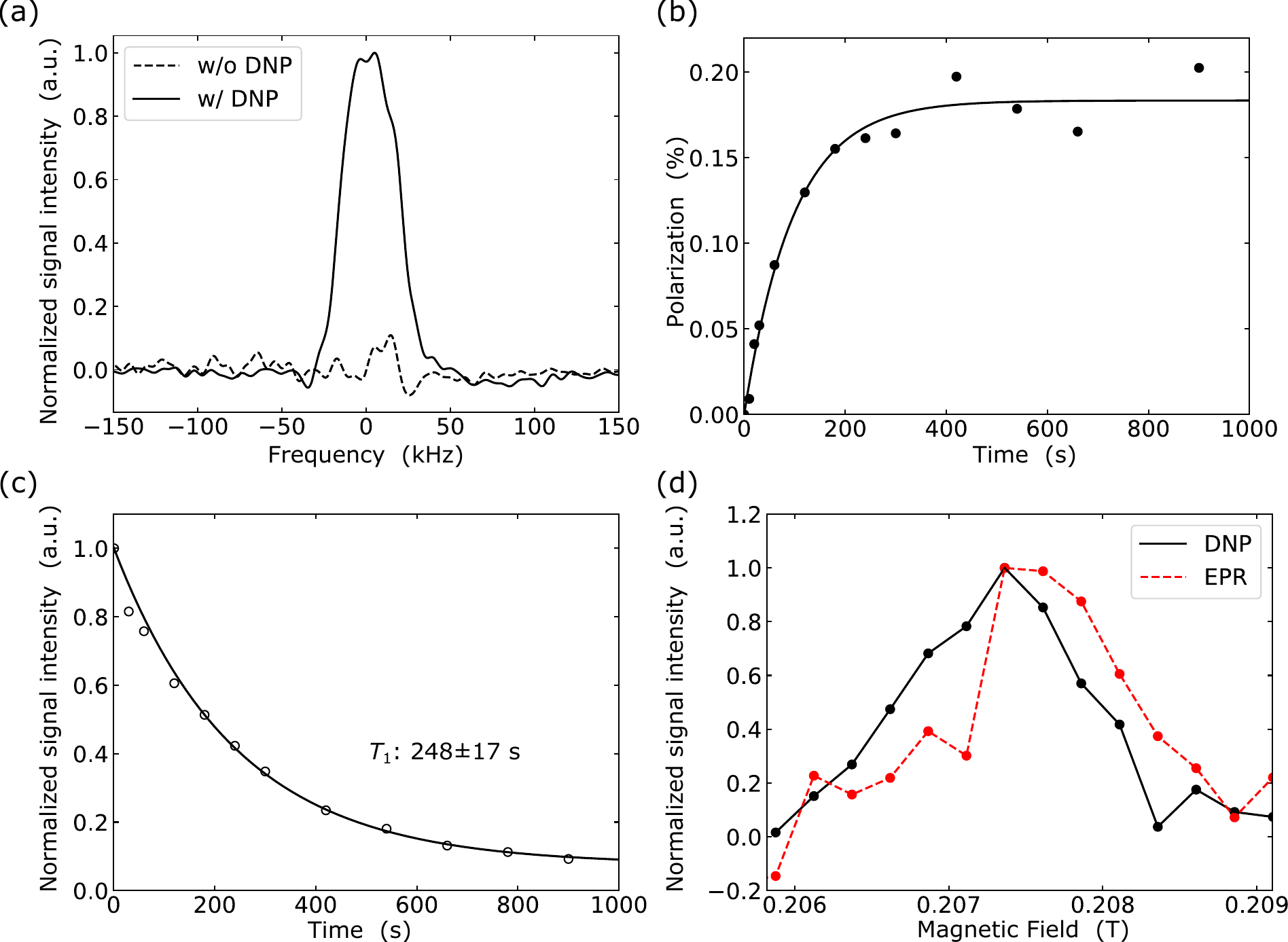}
  \caption{(a) $^1$H NMR spectra of PHPT at a nuclear Larmor frequency $\omega_{0,\mathrm{H}} = 8.74$~MHz with and without DNP for 900~s. The $^1$H polarization $P_{\mathrm{max}}$ of $(0.183\pm0.005)$\% corresponds to a 2600-fold enhancement with respect to thermal equilibrium. (b) Buildup curves of $^1$H polarization for PHPT with ISE repetition rates of 50~Hz. (c) The relaxation curve for the $^1$H NMR signal of PHPT. The $^1$H longitudinal relaxation time $T_1$ was $248\pm17$~s (d) The DNP and EPR field profile of PHPT around 0.207~T. The overlapping spectral profiles confirm that the polarization mechanism is indeed the ISE.}
  \label{fig:DNP_1H_polarization}
\end{figure*}

Adopting these our current-best parameters, we performed dynamic $^1$H polarization by repeating the ISE sequence over the overtone transition at a rate of 50~Hz for 900~s and measured $^1$H NMR in the same magnetic field at the nuclear Larmor frequency $\omega_{\mathrm{0,n}}$ of $2\pi\cdot 8.74$~MHz.
Figure~\ref{fig:DNP_1H_polarization}(a) shows a hyperpolarized $^1$H spectrum obtained in the PHPT sample using a magic-sandwich echo sequence~(MSE)~\cite{rhim1971time}.
The identical $^1$H measurement without overtone triplet-DNP shows no appreciable sign of the NMR signal above the noise level. 
As demonstrated in Fig.~\ref{fig:DNP_1H_polarization}(b), the $^1$H polarization was built up as repeating ISE at a rate of 50~Hz.
The buildup curve fitted with $P(t)=P_{\mathrm{max}}(1-\exp(-t/T_{\mathrm{build}}))$ resulted in the final $^1$H polarization $P_{\mathrm{max}}$ of $(0.183\pm0.005)$\%, corresponding to a 2600-fold enhancement of $^1$H polarization compared to thermal equilibrium.
The time constant $T_{\mathrm{build}}$ of the buildup was $137\pm16$~s.
Figure~\ref{fig:DNP_1H_polarization}(c) shows the profile of $^1$H depolarization after hyperpolarization, from which the time constant $T_1$ of $^1$H longitudinal relaxation was determined to be $240\pm13$~s.
We also performed dynamic $^1$H polarization of pentacene-$d_{14}$-doped $p$-terphenyl~(PDPT) with the same ISE sequence. As shown in Fig.~S4, the maximum $^1$H polarization was $(0.119\pm 0.004)$\%, buildup time constant was and $97\pm11$~s, and longitudinal relaxation time was $248\pm17$~s.
The difference in the attained polarization between PHPT and PDPT arises from the different coupling strength between the protons and the electrons.
In PDPT where the pentacene molecules are fully deuterated, the distance from the electron to the nearest protons situated in the host \textit{p}-terphenyl molecules is longer.
In these measurements, PHPT sample of 0.48~mg and PDPT sample of 0.57~mg were used.

The attainable nuclear spin polarization is determined by several factors, such as the nuclear spin relaxation rate, the densities of the proton spins and the electron spins in the triplet state, the average electron spin polarization, and the exchange probability~\cite{miyanishi2021room, takeda2001dynamic}, which refers to the probability of undergoing exchange of the spin states between the electron and the nuclei in the course of a single ISE sequence.
For both PHPT and PDPT, the density of the electron spins, the average electron spin polarizations, the proton spin density, and the nuclear relaxation times were similar.
The exchange probability depends on the coupling strength between the nuclear and electron spins.
Because the electron spins in the photo-excited triplet state of pentacene are closer to the protons in PHPT than those in PDPT, the exchange probability is the main factor causing difference in the buildup behavior and thereby in the attainable proton polarization. The result that the final polarization was higher for PHPT than for PDPT suggests that the exchange probability was indeed higher for PHPT.

The final $^1$H polarization, 0.183~\%, obtained by the overtone triplet DNP, was lower than the polarization attained (0.75~\%) in our previous work of triplet DNP using the single quantum transition~\cite{kagawa2023triplet}.
Considering that the underlying polarization mechanism is indeed the ISE, which is confirmed by the similarity of the profiles of the magnetic-field dependence of the enhanced $^1$H signal and of the overtone EPR spectrum (Fig.~\ref{fig:DNP_1H_polarization}(d)), a lack of microwave excitation power required to fulfill the Hartmann--Hahn condition of Eq.~(\ref{eq:HH}) is the major factor limiting the current enhancement.
With the on-resonance Rabi nutation frequency of no more than ca.~1.2~MHz (Fig.~\ref{fig:DNP_optimisation}(c)) and several times higher $^{1}$H Larmor frequency $\omega_\mathrm{0,H} = 8.74$~MHz, the Hartmann-Hahn condition needs to be satisfied with a rather large offset magnetic field, which can reduce the secular term of the hyperfine interaction that contribute to exchange of mutual spin flip and thereby deteriorate the exchange probability during ISE.
Thus, a higher overtone Rabi nutation frequency and a larger DNP enhancement would improve the attainable polarization, and this required a microwave amplifier with a higher output power.

We also tried to perform hyperpolarization of the $^{13}$C spins in microdiamonds with NV$^{-}$ centers in a similar setup with a $^{13}$C NMR frequency of 2.12~MHz, but no $^{13}$C signal was observed.
Presumably, the low natural abundance (ca.~1~\%) of $^{13}$C and a rather small gyromagnetic ratio compared to $^1$H hindered both efficient electron-to-$^{13}$C polarization transfer and spin diffusion among the $^{13}$C spins. 
The relatively low polarization of NV$^-$ centers also contributes to the difficulty in observing DNP-enhanced $^{13}$C NMR signals.

From the overtone-DNP perspective, it is important to note that the nuclear polarization is distributed over nearly all electron spin and microcrystal orientations. This is in contrast to triplet-DNP on the $|\Delta M_S| = 1$ transition, where only certain electron spins or crystallite orientations participate in the polarization process~\cite{miyanishi2021room}.
In addition, the fraction of the overtone states $|\pm 1 \rangle$ affects the rate at which proton polarization is built up.
Of the pentacene molecules photo-excited to the triplet state, those populated onto the $|0 \rangle$ state are irrelevant.
Thus, the density of the sources of polarization is lower than the concentration of those pentacene molecules undergoing intersystem crossing to the triplet state upon photo-excitation.
If the fraction of the overtone state is rather small, it may seem that the source density would become exceedingly small.
However, it is important to note that the state in which each pentacene molecule settles depends on the probability.
In triplet-DNP experiments where the cycle of photo-excitation and ISE is repeated multiple times, every pentacene molecule has a finite chance to be in the active states, and therefore contributes to polarization transfer.
It is not the source density but the frequency of repetition of the DNP process that is affected by the small fraction of the overtone states.
This technical issue could be addressed by introducing laser equipment capable of pulsing at a higher rate.

\section{Conclusions and Outlook}
In this work, we demonstrated the first successful acquisition of the overtone Rabi oscillations of the electron spins in the optically polarized triplet state of pentancene and NV$^{-}$ centers in powder. In addition, we implemented triplet-DNP exploiting the hyper-overtone-polarization, building up the $^{1}$H polarization at room temperature in ca.~0.2~T up to $0.183\pm0.005$\%, corresponding to a 2600-fold NMR signal enhancement with respect to thermal polarization.
In such a high-field regime, where the Zeeman interaction is dominant over the zero-field splitting interaction, the single-quantum spectrum is too broad to be excited entirely in a single sweep over resonance and therefore cannot be fully exploited as a source of nuclear hyperpolarization.
In this respect, the overtone transition has an advantage.
The residual broadening of the resonance condition is second-order and is scaled down to such an extent that it can be conveniently swept over with the conventional ISE scheme.
We also established analytical formulas for evaluating the shift and the residual broadening of the overtone transition for a given combination of the ZFS interaction and the magnetic-field strength in the axially symmetric case.
In addition, we derived the analytical expressions for the nutation spectrum in the case of orthogonal and parallel orientations of the microwave field to the external magnetic field.
These formulae help understanding the nature of the overtone transition in terms of the resonance position and its distribution in randomly oriented polycrystalline systems.
In the presence of finite asymmetry, one may resort to numerical evaluations.
The current limitation of the finally attained nuclear polarization is mainly due to technical factors (microwave power and laser repetition rate) and could be improved in future.

The overtone DNP could become advantageous for systems with a larger value of $\varepsilon=\omega_{\mathrm{ZFS}}/\omega_{\mathrm{e}}$.
On the one hand, this will improve the efficiency of overtone DNP by an increased Rabi nutation frequency, which scales with $\varepsilon=\omega_{\mathrm{ZFS}}/\omega_{\mathrm{e}}$. On the other hand, the efficiency of the single-quantum DNP could decrease, as the powder linewidth could become too broad for the larger ZFS. Therefore, it would be interesting to test a system with ZFS parameters similar to those of NV$^-$ centers in diamond, but having $^1$H spins in the vicinity of the polarized $S=1$ electrons. This could be achieved chemically by modifying the pentacene ring, e.g., 5,12-diazatetracene with a ZFS-parameter of $D \approx 2\pi \cdot 1.6$~GHz \cite{kouno2019} or reducing the ring size, e.g., a benzene molecule, leading to a strong increase with $D \approx 2\pi \cdot 4.6$~GHz \cite{mcglynnMolecularSpectroscopyTriplet1969}.

\begin{acknowledgments}
This work was supported by the Ministry of Education, Culture, Sports, Science and Technology Quantum Leap Flagship Program (MEXT Q-LEAP; Grant no. JPMXS0120330644), a Kakenhi Grant-in-Aid (Grant no. 23K13775) from the Japan Society for the Promotion of Science (JSPS), World Premier International Research Center Initiative (WPI), MEXT, Japan, and JST Adopting Sustainable Partnerships for Innovative Research Ecosystem (ASPIRE) Program (Grant no. JPMJAP2319.
We thank Izuru Ohki, Hiroshi Abe, and Norikazu Mizuochi for supplying microdiamonds containing NV$^-$ centers. TFS thanks Alexander I. Shames (Ben-Gurion University of the Negev) for his introduction to the half-field transition and suggesting the half-field DNP experiment. Writefull in Overleaf was used to improve English phrasing.

\end{acknowledgments}

\clearpage 
\onecolumngrid
\appendix 
\input{arxiv_supplementary_2}

\end{document}

%% file: arxiv_supplementary_2.tex
\renewcommand{\thefigure}{S\arabic{figure}}
\renewcommand{\theequation}{S\arabic{equation}}

\title{Supplemental Material for Overtone Rabi oscillation of optically polarized triplet electron spins and nuclear hyperpolarization in powder}

\author{Koichiro Miyanishi}
\email{miyanishi.koichiro.qiqb@osaka-u.ac.jp; Present address: Qubitcore Inc., OIST Inovation core2 OIC2 207 1919-1 Tancha, Onna-son, Kunigami-gun Okinawa 904-0495, Japan}
\affiliation{Center for Quantum Information and Quantum Biology, The University of Osaka, 1-2 Machikaneyama, Toyonaka, Osaka 560-0043 Japan}

\author{Takuya F. Segawa}
\affiliation{Institute of Molecular Physical Science / Laboratory of Physical Chemistry, Department of Chemistry and Applied Biosciences, ETH Zurich, 8093 Zurich, Switzerland}

\author{Makoto Negoro}
\affiliation{Center for Quantum Information and Quantum Biology, The University of Osaka, 1-2 Machikaneyama, Toyonaka, Osaka 560-0043 Japan}
\affiliation{Institute for Quantum Life Science (iQLS), National Institutes for Quantum Science and Technology (QST), Chiba 263-8555, Japan}
\affiliation{Premium Research Institute for Human Metaverse Medicine, The University of Osaka, Suita, Osaka 565-0871, Japan}

\author{Akinori Kagawa}
\affiliation{Center for Quantum Information and Quantum Biology, The University of Osaka, 1-2 Machikaneyama, Toyonaka, Osaka 560-0043 Japan}
\affiliation{Premium Research Institute for Human Metaverse Medicine, The University of Osaka, Suita, Osaka 565-0871, Japan}

\author{Kazuyuki Takeda}
\email{takezo@kuchem.kyoto-u.ac.jp}
\affiliation{ Division of Chemistry, Graduate School of Science, Kyoto University, Kyoto 606-8502, Japan}

\maketitle

\newpage

\section{Derivation of Eq.~(13)-(14)}
The ZFS interaction $H_{\mathrm{ZFS}}$ can be expressed using rank-2 spherical tensor spin operators $T_{2,-k}$ and rank-2 spherical harmonics $A_{2,k}(\Omega)$ as
\begin{align}
  H_{\mathrm{ZFS}} = \sum_{k=-2}^{2}(-1)^{k} A_{2,k}(\Omega)T_{2,-k},
\end{align}
where $\Omega=(\alpha,\beta,\gamma)$ are the Euler angles describing the orientation of the principal axis system (PAS) of the ZFS tensor.

Following a formalism of overtone NMR for $^{14}$N (spin 1) ~\cite{tycko1987overtone,marinelli1999density} where the quadrupolar interaction comes into play, we express the ZFS interaction as
\begin{align}
    H_{\mathrm{ZFS}}&=\omega_{\mathrm{ZFS}} \left[
    \sqrt{6} h T_{2,0}+ fT_{2,1}-f^{\ast}T_{2,-1} \right. \nonumber \\
                    & \left. \ \ \ \ \ \ \ \ \ \ +gT_{2,2}+g^{\ast}T_{2,-2} \right],  \\
    h'(\alpha,\beta)&=\frac{1}{2} [(3\cos^2\beta-1) +\eta\cos2\alpha\sin^2\beta ],  \\
    f(\alpha,\beta,\gamma)&= e^{i\gamma}[3\sin\beta \cos\beta \nonumber \\
                &\ \ \ \ \  -\eta(\cos 2\alpha \sin\beta \cos\beta + i \sin 2\alpha \sin \beta) ],  \\
    g(\alpha,\beta,\gamma)&= e^{2i\gamma}\frac{1}{2}[3\sin^2\beta \nonumber \\
                &\ +\eta(\cos 2\alpha (1+\cos^2\beta) + 2i\sin2\alpha\cos\beta) ], \\
    \omega_{\mathrm{ZFS}} &= \frac{D}{3}, \\
    \eta &= \frac{3E}{D}.
\end{align}
Here, $D$ and $E$ are the ZFS parameters~\cite{weil2007electron}. 
In the case of the electron spin in the photo-excited triplet state of pentacene doped in $p$-terphenyl, $D$ and $E$ are $2\pi\cdot 1395.57$~MHz and $2\pi\cdot 53.35$~MHz~\cite{yang2000zero}, whereas for the NV$^-$ center in diamond, $D$ and $E$ are $2\pi\cdot 2870$~MHz and $\approx0$~\cite{doherty2013nitrogen}.
In a magnetic field of the order of 0.2~T used in this work for the overtone experiments, the electron Larmor frequency $\omega_e=2\pi\cdot 5.8$~GHz is much higher than $\omega_{\mathrm{ZFS}}\simeq 2\pi\cdot 465$~MHz for pentacene and $\omega_{\mathrm{ZFS}}\simeq 2\pi\cdot 957$~MHz for NV$^-$ centers.
To appreciate that $\omega_{\mathrm{ZFS}}$ is considerably lower than $\omega_{e}$, we introduce
\begin{align}
    \varepsilon \equiv \frac{\omega_{\mathrm{ZFS}}}{\omega_{e}} \ll 1.
\end{align}

The matrix representation of the static part of the Hamiltonian is
\begin{align}
H_{\mathrm{static,e}} &= 
H_{\mathrm{Zeeman,e}}+H_{\mathrm{ZFS}}  \nonumber \\
&=
\omega_{e}
\begin{bmatrix}
1+\varepsilon h' & -\frac{1}{\sqrt{2}}\varepsilon f & \varepsilon g \\
-\frac{1}{\sqrt{2}}\varepsilon f^{\ast} & -2\varepsilon h'  & \frac{1}{\sqrt{2}}\varepsilon f \\
\varepsilon g^{\ast} & \frac{1}{\sqrt{2}}\varepsilon f^{\ast} & -1 + \varepsilon h' 
\end{bmatrix}.
\label{eq:static_hamiltonian}
\end{align}
Assuming that the ZFS interaction can be taken as a perturbation to the dominant Zeeman interaction, we apply the Schrieffer–Wolff (SW) transformation~\cite{luttinger1955motion,schrieffer1966relation}, namely, such unitary transformation that diagonalizes the Hamiltonian in first order, leaving only second-order off-diagonal elements:
\begin{align}
H_{\mathrm{static,e}}^{\mathrm{SW}}&=U^{\dagger}H_{\mathrm{static,e}}U=e^{T}He^{-T}.
\end{align}
Here, $T$ is an anti-hermitian operator satisfying $H_{\mathrm{static},1}=[H_{\mathrm{static},0},T]$, where $H_{\mathrm{static},0(1)}$ is the diagonal (off-diagonal) part of the static Hamiltonian $H_{\mathrm{static,e}}$. The anti-hermitian operator $T$ is given as:
\begin{align}
T= \varepsilon
\left[\begin{matrix}
0 & - \frac{1}{1 + 3 \varepsilon h} \frac{f}{\sqrt{2}} & \frac{g}{2} \\
\frac{1}{1 + 3 \varepsilon h} \frac{f^{\ast}}{\sqrt{2}}  & 0 & \frac{1}{1 - 3 \varepsilon h} \frac{f}{\sqrt{2}}  \\
- \frac{g^{\ast}}{2}  & -\frac{1}{1 - 3 \varepsilon h} \frac{f^{\ast}}{\sqrt{2}}   & 0
\end{matrix}\right].
\end{align}
By the SW transformation, the static Hamiltonian becomes
\begin{align}
  H_{\mathrm{static,e}}^{\mathrm{SW}}&= H_{\mathrm{static},0} 
  + \left[T,H_{\mathrm{static},1}\right]+\mathcal{O}\left(T^2\right) \nonumber \\
&= \left[\begin{matrix}
 \omega_{e} + \omega_{\mathrm{ZFS}} h' & 0 & 0 \\ 0 & -2 \omega_{\mathrm{ZFS}} h' & 0 \\ 0 & 0 & -\omega_{e} + \omega_{\mathrm{ZFS}} h'
\end{matrix}\right] \\ 
&+ \varepsilon \omega_{\mathrm{ZFS}}\left[\begin{matrix}
f f^{\ast} + g g^{\ast}
& \frac{3\sqrt{2}}{4} f^{\ast} g 
& 0 \\
\frac{3\sqrt{2}}{4} f g^{\ast} 
& 0
& \frac{3\sqrt{2}}{4} f^{\ast} g \\
0
& \frac{3\sqrt{2}}{4} f g^{\ast} 
& - f f^{\ast} - g g^{\ast}
\end{matrix}\right] \\
&+\mathcal{O}\left(\varepsilon^2\right).
\end{align}

Similarly, $H_{\mathrm{MW}}$ is SW transformed as
\begin{align}
H_{\mathrm{MW}}^{\mathrm{SW}}&= 2\omega_{1} \left(\sin\chi S_x^{\mathrm{SW}}+\cos\chi S_z^{\mathrm{SW}} \right)\cos (\omega_{\mathrm{MW}}t),
\end{align}
with
\begin{align}
  S_x^{\mathrm{SW}} &= e^{T} S_{x} e^{-T} = S_{x} + [T,S_{x}] + \mathcal{O}(T^{2})\nonumber\\
&=\frac{1}{\sqrt{2}}\left[\begin{matrix}
0 & 1 & 0 \\ 1 & 0 & 1 \\ 0 & 1 & 0
\end{matrix}\right] \nonumber \\
&+ \varepsilon \left[ \begin{matrix}
 -\Re{f} & \frac{\sqrt{2}}{2} g & -f \\
 \frac{\sqrt{2}}{2} g^{\ast} & 2\Re{f} & -\frac{\sqrt{2}}{2} g \\
 -f^{\ast} & -\frac{\sqrt{2}}{2} g^{\ast} & -\Re{f}
\end{matrix}\right] \nonumber \\
&+\mathcal{O}(\varepsilon^{2}),
\end{align}
and
\begin{align}
  S_z^{\mathrm{SW}} &= e^{T}S_{z}e^{-T} = S_{z} + [T,S_{z}] + \mathcal{O}(T^{2}) \nonumber\\
&= \left[ \begin{matrix}
  1 & 0 & 0 \\ 0 & 0 & 0 \\ 0 & 0 & -1 
\end{matrix}\right] \nonumber \\
&+ \varepsilon \left[ \begin{matrix}
    0 & \frac{1}{\sqrt{2}} f & -g \\
    \frac{1}{\sqrt{2}} f^{\ast} & 0 & -\frac{1}{\sqrt{2}} f \\
    -g^{\ast} & -\frac{1}{\sqrt{2}} f^{\ast} & 0
\end{matrix}\right] \nonumber \\
&+ \mathcal{O}(\varepsilon^{2}).
\end{align}

Next, we consider a reference frame rotating around the $z$ axis at frequency $\omega_{\mathrm{MW}}/2$.
The net Hamiltonian $H$ in this frame is given by
\begin{align}
    H = R^{\dagger}(H_{\mathrm{static,e}}^{\mathrm{SW}} + H_{\mathrm{MW}}^{\mathrm{SW}})R - \frac{1}{2}\omega_{\mathrm{MW}} S_{z},
\end{align}
with
\begin{align}
    R = \exp\left[-i \frac{\omega_{\mathrm{MW}}}{2}t S_{z} \right].
\end{align}
Keeping $\omega_{\mathrm{e}} \approx \omega_{\mathrm{MW}}/2$ in mind, we introduce an offset frequency $\Delta\omega$ defined as 
\begin{align}
    \Delta \omega = \omega_{e} - \frac{\omega_{\mathrm{MW}}}{2}.
\end{align}
Discarding terms oscillating at frequencies of $\omega_{\mathrm{MW}}/2$ and higher, we obtain
\begin{align}
  H = \left[ \begin{matrix}
      \Delta \omega + \varepsilon \omega_{\mathrm{ZFS}} ( f f^{\ast} + g g^{\ast})  & 0 & -\varepsilon \omega_{1} (f \sin\chi + g \cos\chi) \\
      0 & -3\omega_{\mathrm{ZFS}}h' & 0 \\
      -\varepsilon \omega_{1} (f^{\ast} \sin\chi + g^{\ast}\cos\chi) & 0 &  -\Delta \omega - \varepsilon \omega_{\mathrm{ZFS}} ( f f^{\ast} + g g^{\ast})
  \end{matrix}\right] 
  + \omega_{\mathrm{ZFS}}h' \cdot \mathbf{1}
\end{align}

The resonance condition for the overtone transition between the tilted $\ket{+1}$ and $\ket{-1}$ eigenstates is obtained by requesting $(1,1)$ or $(3,3)$ element of $H$ vanishes:
\begin{align}
    \omega_{\mathrm{MW}} = 2 \omega_{e} + 2 \varepsilon \omega_{\mathrm{ZFS}} ( f f^{\ast} + g g^{\ast}).
    \label{eq:resonance-cond}
\end{align}
Eq.~(\ref{eq:resonance-cond}) indicates that the resonance frequency of the overtone transition is not exactly two times the electron Larmor frequency $\omega_{e}$, but has a shift $2 \varepsilon \omega_{\mathrm{ZFS}} h$, where $h$ is
\begin{align}
h = ( f f^{\ast} + g g^{\ast}).
    \label{eq:h2}
\end{align}

Eqs.~(\ref{eq:resonance-cond})-(\ref{eq:h2}) lead to Eq.~(13).
The nutation frequency $\omega_{\mathrm{nut}}$ is given by the magnitude of the $(1,3)$ or $(3,1)$ element of $H$, that is, Eq.~(14).

\section{Lineshape of overtone resonance}
The materials dealt with in the present work come with axially symmetric or nearly axially symmetric ZFS tensor.
Here, we derive a lineshape function $I(\omega')$ of the overtone spectrum in the limit of $\eta \rightarrow 0$, where $\omega' \equiv \omega - 2\omega_{e}$ is an angular frequency referenced from the unperturbed overtone frequency $2\omega_{e}$.
Now the anisotropic shift is reduced to
\begin{align}
    \omega'=\frac{9}{2} \varepsilon\omega_{\mathrm{ZFS}}\sin^{2}\beta (4\cos^{2}\beta+\sin^{2}\beta).
\end{align}
The lineshape function $I(\omega')$ satisfies
\begin{align}
  1 = \int_{-\infty}^{\infty} d\omega' I(\omega') 
    = \frac{1}{4\pi} \int_{0}^{2\pi} d\gamma \int_{0}^{\pi} \sin\beta d\beta.
    \label{eq:int1}
\end{align}
In the present case, the integral over $\gamma$ is simply $2\pi$ and the $\beta$-dependence is symmetric about $\beta=\pi/2$, so that
\begin{align}
    \int d\omega I(\omega) = \frac{1}{4\pi} \cdot 2\pi \cdot 2\int_{0}^{\pi/2} \sin\beta d\beta = \int_{0}^{\pi/2} \sin\beta d\beta.
    \label{eq:int2}
\end{align}
To obtain $I(\omega)$, we aim to rewrite the integral on the right side of the above equation in terms of $d\omega$ so that we can set the integrands on both sides equal to each other.

We introduce a variable $x=-\cos\beta$. Then, $x$ varies from $-1$ to $0$ as $\beta$ does from $0$ to $\pi/2$, so that 
\begin{align}
    \int_{0}^{\pi/2} \sin\beta d\beta = \int_{-1}^{0} dx,
    \label{eq:int3}
\end{align}
and we have
\begin{align}
    \omega' = \frac{9}{2} \varepsilon \omega_{\mathrm{ZFS}} (1-x^{2}) (1 + 3 x^{2}),
    \label{eq:omega_x}
\end{align}
and
\begin{align}
    d\omega'= 18 \varepsilon \omega_{\mathrm{ZFS}} x (1 - 3 x^{2}) dx,
    \label{eq:domega-dx}
\end{align}
or
\begin{align}
    dx = \frac{1}{18\varepsilon\omega_{\mathrm{ZFS}}} x^{-1} (1-3 x^{2})^{-1} d\omega'
\end{align}
Noting that $-1 \leq x \leq 0$, we see $\omega' \geq 0$, $\omega'(x=-1)=0$, $\omega'(x=0) = \frac{9}{2}\varepsilon\omega_{\mathrm{ZFS}}$, and 
$\omega'$ is always equal to or greater than 0.
From Eq.~(\ref{eq:domega-dx}), we also see that the slope of $\omega'(x)$ is zero at $x=0$ and $x=-1/\sqrt{3}$, and the maximum value is $\omega'(x=-1/\sqrt{3}) = 6 \varepsilon \omega_{\mathrm{ZFS}}$.

For $0\leq \omega' \leq \frac{9}{2}\varepsilon\omega_{\mathrm{ZFS}}$ ($-1 \leq x \leq -\sqrt{\frac{2}{3}}$), there exists a single solution $x_{1}$ for Eq.~(\ref{eq:omega_x}),
\begin{align}
    x_{1} = -\frac{1}{\sqrt{3}}\sqrt{1+2\sqrt{1-\frac{\omega'}{6\varepsilon\omega_{\mathrm{ZFS}}}}}.
\end{align}
For $\frac{9}{2}\varepsilon\omega_{\mathrm{ZFS}} \leq \omega' \leq 6\varepsilon\omega_{\mathrm{ZFS}}$, we find a pair of solutions, one of which is $x_{1} ~ (-\sqrt{\frac{2}{3}} \leq x \leq -1/\sqrt{3})$, and another, $x_{2} ~ (-1/\sqrt{3} \leq x \leq 0)$, is given by:
\begin{align}
        x_{2} = -\frac{1}{\sqrt{3}}\sqrt{1-2\sqrt{1-\frac{\omega'}{6\varepsilon\omega_{\mathrm{ZFS}}}}}.
\end{align}
It follows that
\begin{align}
    \int_{-1}^{0} dx &= \int_{-1}^{-\sqrt{2/3}} dx
                      + \int_{-\sqrt{2/3}}^{-1/\sqrt{3}} dx
                      + \int_{-1/\sqrt{3}}^{0} dx \\
    &= \frac{1}{18\varepsilon\omega_{\mathrm{ZFS}}} \int_{0}^{\frac{9}{2}\varepsilon\omega_{\mathrm{ZFS}}} d\omega' x_{1}^{-1} (1-3x_{1}^{2})^{-1} \nonumber \\
    &+ \frac{1}{18\varepsilon\omega_{\mathrm{ZFS}}} \int_{\frac{9}{2}\varepsilon\omega_{\mathrm{ZFS}}}^{6\varepsilon\omega_{\mathrm{ZFS}}} d\omega' x_{1}^{-1} (1-3x_{1}^{2})^{-1} \nonumber \\
    &+ \frac{1}{18\varepsilon\omega_{\mathrm{ZFS}}} \int_{6\varepsilon\omega_{\mathrm{ZFS}}}^{\frac{9}{2}\varepsilon\omega_{\mathrm{ZFS}}} d\omega' x_{2}^{-1} (1-3x_{2}^{2})^{-1}
    \label{eq:int4}
\end{align}
From Eqs.~(\ref{eq:int2})-(\ref{eq:int4}), we obtain the lineshape function $I(\omega')$ to be
\begin{align}
    I(\omega') = \frac{1}{18}\sqrt{\frac{3}{2}} \frac{1}{\varepsilon\omega_{\mathrm{ZFS}}} \left[1-\frac{\omega'}{6\varepsilon\omega_{\mathrm{ZFS}}} \right]^{-\frac{1}{2}} \left[ 1+2\sqrt{1-\frac{\omega'}{6\varepsilon\omega_{\mathrm{ZFS}}}}  \right]^{-\frac{1}{2}},
\end{align}
for $0 \leq \omega' \leq \frac{9}{2}\varepsilon\omega_{\mathrm{ZFS}}$ and
\begin{align}
    I(\omega') = \frac{1}{18}\sqrt{\frac{3}{2}} \frac{1}{\varepsilon\omega_{\mathrm{ZFS}}} \left[1-\frac{\omega'}{6\varepsilon\omega_{\mathrm{ZFS}}} \right]^{-\frac{1}{2}} 
    \left\{
       \left[ 1+2\sqrt{1-\frac{\omega'}{6\varepsilon\omega_{\mathrm{ZFS}}}}  \right]^{-\frac{1}{2}}
       +
       \left[ 1-2\sqrt{1-\frac{\omega'}{6\varepsilon\omega_{\mathrm{ZFS}}}}  \right]^{-\frac{1}{2}} 
    \right\},
\end{align}
for $\frac{9}{2}\varepsilon\omega_{\mathrm{ZFS}} \leq \omega' \leq 6\varepsilon\omega_{\mathrm{ZFS}}$. Eqs.~(15)-(16) in the main text are alternative expressions using $\omega'=\omega-2\omega_{e}$ and $\varepsilon = \omega_{\mathrm{ZFS}}/\omega_{e}$.

\begin{figure}[h]
    \includegraphics[width=0.85\textwidth]{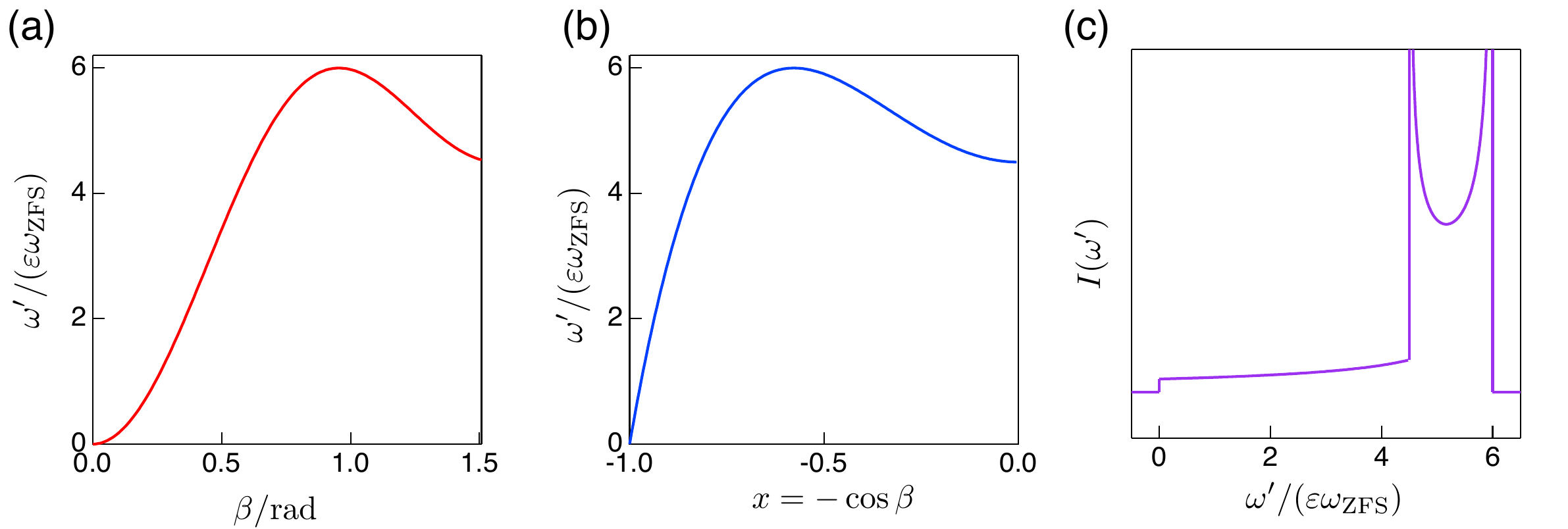}
    \caption{(a) $\beta$ dependence of the overtone transition frequency $\omega'$ for axially symmetric ZFS tensor. (b) $x=-\cos\beta$ dependence (c) A lineshape function $I(\omega')$}
    \label{fig:ot-lineshape}
\end{figure}

\section{Overtone nutation spectrum}
The function representing the distribution of the nutation frequency in randomly oriented ensemble system can be derived in a similar way.
Here, we deal with the case in which the microwave field is normal (parallel) to the static field, so that $\chi=\pi/2$ ($\chi=0$) in Eq.~(14).

\subsection{Microwave field $B_1$ normal to the static field $B_0$ ($\chi=\pi/2$)}
The nutation frequency $\omega_{\mathrm{nut}}$ is given by
\begin{align}
    \omega_{\mathrm{nut}} = \frac{3}{2}\varepsilon\omega_{1}\sin 2\beta.
    \label{eq:omega-nut-perp-beta}
\end{align}
With $\beta$ running from $0$ to $\pi$, $\omega_{\mathrm{nut}}$ can take values between $0$ and $\frac{3}{2}\varepsilon\omega_{1}$.
The distribution function $I_{\perp}(\omega)$ accumulates up to unity when integrated over this range.
\begin{align}
    \int_{0}^{\frac{3}{2}\varepsilon\omega_{1}} d\omega_{\mathrm{nut}} I_{\perp}(\omega_{\mathrm{nut}})
    &=
    \frac{1}{4\pi}\int_{0}^{2\pi} d\gamma \int_{\pi}^{0} \sin\beta d\beta \\
    &= \int_{0}^{\pi/2} \sin\beta d\beta \\
    &= \int_{-1}^{0} dx.
    \label{eq:int40}
\end{align}
Here we introduced a variable $x=-\cos\beta$ again (see the previous section).

We rewrite the nutation frequency $\omega_{\mathrm{nut}}$ in Eq.~(\ref{eq:omega-nut-perp-beta}) in terms of $x$ as
\begin{align}
    \omega_{\mathrm{nut}} = -3 \varepsilon \omega_{1} x \sqrt{1-x^{2}},
\end{align}
so that
\begin{align}
    \frac{d\omega_{\mathrm{nut}}}{dx} &= -3\varepsilon\omega_{1}
    \left[ 
      \sqrt{1-x^{2}}-\frac{x^{2}}{\sqrt{1-x^{2}}}
    \right] \\
    &= -3\varepsilon\omega_{1} \frac{1-2x^{2}}{\sqrt{1-x^{2}}},
\end{align}
or
\begin{align}
    dx = \frac{1}{3\varepsilon\omega_{1}}
         \frac{\sqrt{1-x^{2}}}{2x^{2}-1}
         d\omega_{\mathrm{nut}}.
\end{align}

We also obtain $x$, or $x^{2}$, as a function of $\omega_{\mathrm{nut}}$ to be
\begin{align}
    x_{\pm}^{2} &= \frac{1}{2} \left[ 
      1 \pm \sqrt{1-4 \zeta^{2} }
    \right], \\
       \zeta &\equiv \frac{\omega_{\mathrm{nut}}}{3\varepsilon\omega_{1}}.
\end{align}
Here, the plus sign is taken for $-1 \le x \le -1/\sqrt{2}$, while the sign should be minus for $-1/\sqrt{2} \le x \le 0$.
Accordingly, we divide the integral Eq.~(\ref{eq:int40}) into two parts, i.e., $\int_{-1}^{-1/\sqrt{2}} dx + \int_{-1/\sqrt{2}}^{0} dx$.
When $x$ goes from $-1$ to $-\frac{1}{\sqrt{2}}$, $\omega_{\mathrm{nut}}$ changes from $0$ to $\frac{3}{2}\varepsilon\omega_{1}$.
Conversely, $\omega_{\mathrm{nut}}$ deceases from $\frac{3}{2}\varepsilon\omega_{1}$ down to $0$ as $x$ sweeps over the range between $-\frac{1}{\sqrt{2}}$ and $0$.
Thus,
\begin{align}
    \int_{0}^{\frac{3}{2}\varepsilon\omega_{1}} d\omega_{\mathrm{nut}} I_{\perp}(\omega_{\mathrm{nut}})
    =
    \frac{1}{3\varepsilon\omega_{1}}
    \int_{0}^{\frac{3}{2}\varepsilon\omega_{1}} d\omega_{\mathrm{nut}} 
         \frac{\sqrt{1-x_{+}^{2}}}{2x_{+}^{2}-1}
    - 
    \frac{1}{3\varepsilon\omega_{1}}
    \int_{0}^{\frac{3}{2}\varepsilon\omega_{1}} d\omega_{\mathrm{nut}}
         \frac{\sqrt{1-x_{-}^{2}}}{2x_{-}^{2}-1},
\end{align}
and we finally arrive at
\begin{align}
    I_{\perp}(\omega_{\mathrm{nut}}) =
    \frac{1}{3\sqrt{2}\varepsilon\omega_{1}}
    \left[ 1 - 4 \left( \frac{\omega_{\mathrm{nut}}}{3\varepsilon\omega_{1}} \right)^{2}\right]^{-\frac{1}{2}} 
    \left[ 
    \sqrt{1-\sqrt{1-4\left(\frac{\omega_{\mathrm{nut}}}{3\varepsilon\omega_{1}} \right)^{2}}} 
    +
    \sqrt{1+\sqrt{1-4\left(\frac{\omega_{\mathrm{nut}}}{3\varepsilon\omega_{1}} \right)^{2}}} 
    \right].
\end{align}

\subsection{Microwave field $B_1$ parallel to the static field $B_0$ ($\chi=0$)}
By inspecting the nutation frequency
\begin{align}
    \omega_{\mathrm{nut}} = \frac{3}{2}\varepsilon\omega_{1}\sin^{2}\beta = \frac{3}{2}\varepsilon\omega_{1}\sin^{2}(1-x^{2}),
\end{align}
we see that the distribution function is well defined in a range $0 \leq \omega_{\mathrm{nut}} \leq \frac{3}{2}\varepsilon\omega_{1}$, the maximum being half that in the case of $\chi=\pi/2$.
Noting that $x<0$, we see
\begin{align}
    x = - \sqrt{1-\frac{2\omega_{\mathrm{nut}}}{3\varepsilon\omega_{1}}},
\end{align}
and
\begin{align}
  \frac{d\omega_{\mathrm{nut}}}{dx} = -3\varepsilon\omega_{1} x.    
\end{align}
Through the same process, we obtain
\begin{align}
    I(\omega_{\mathrm{nut}}) = \frac{1}{3\varepsilon\omega_{1}}
    \left[ 
    1-\frac{2\omega_{\mathrm{nut}}}{3\varepsilon\omega_{1}}
    \right]^{-\frac{1}{2}}
\end{align}

\section{Experimental apparatus for EPR and DNP experiments}
\begin{figure}[h]
    \includegraphics[width=90mm]{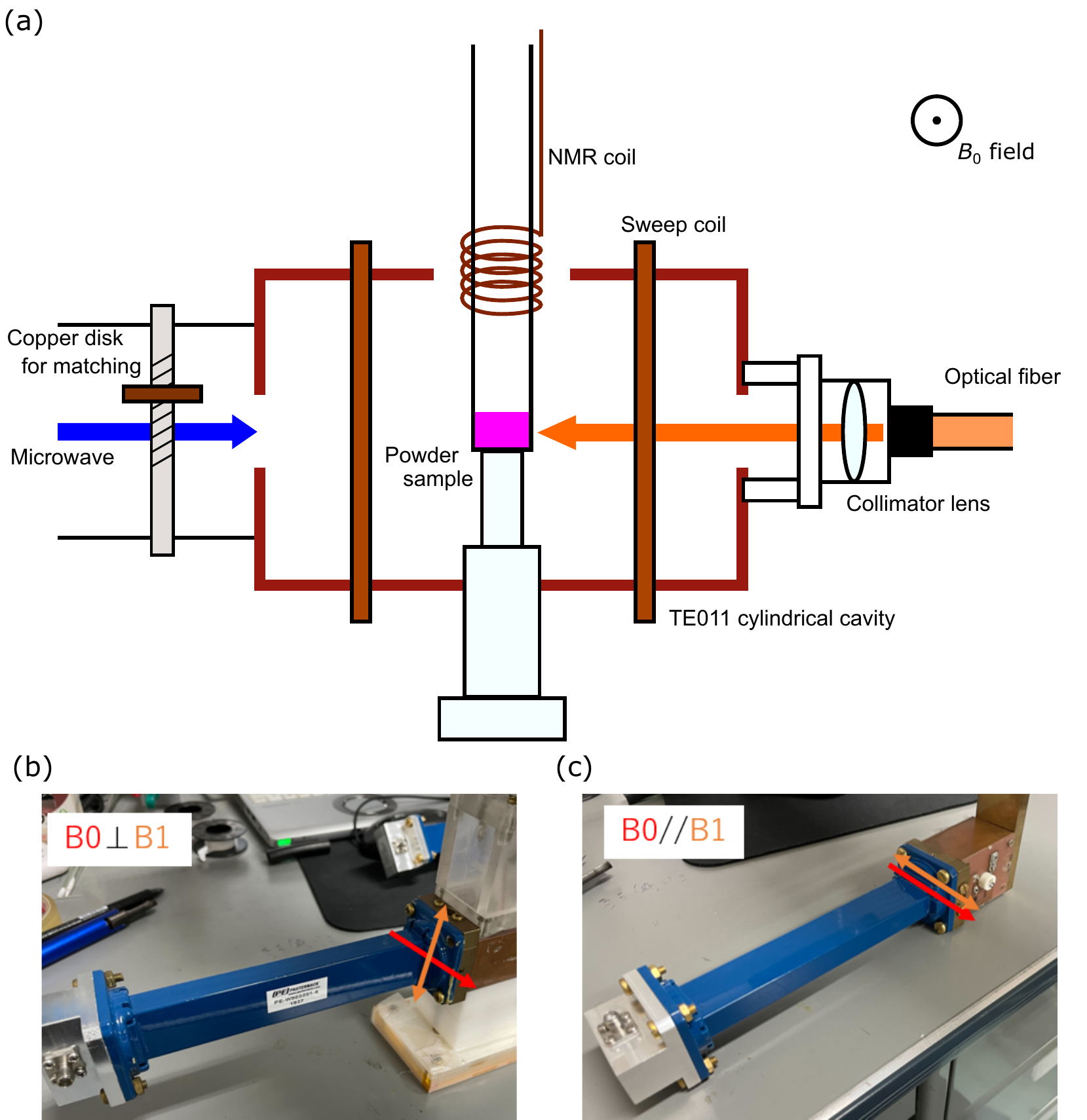}
    \caption{(a) The drawing of the experimental apparatus used for EPR and DNP measurements. (b) Photo of the cylindrical cavity and the rectangular waveguide for EPR measurement with $B_0 \perp B_1$. (c) Photo of the cylindrical cavity and the rectangular waveguide for EPR measurement with $B_0 \parallel B_1$.}
    \label{fig:Experimental_apparatus}
\end{figure}

Figure~\ref{fig:Experimental_apparatus}(a) shows a detailed drawing of the experimental apparatus for EPR and DNP experiments. This cylindrical cavity was set in an electromagnet with a static magnetic field from 0.19 to 0.5~T. Photos of the cylindrical cavity were shown in Fig.~\ref{fig:Experimental_apparatus}(b) and (c). The setting of Fig.~\ref{fig:Experimental_apparatus}(b) was used for the EPR and DNP measurements in the manuscript. The setting of Fig.~\ref{fig:Experimental_apparatus}(c) was used to perform EPR measurement with $B_0 \parallel B_1$ using triplet electrons in NV$^-$ centers in diamonds.

\newpage

\section{Molecular structures of the sample}
\begin{figure}[h]
    \centering
    \includegraphics[width=0.4\linewidth]{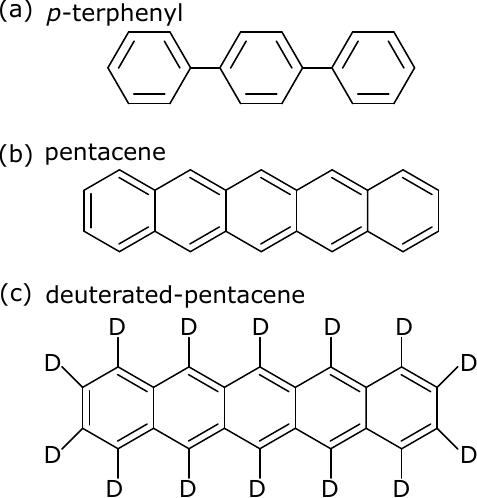}
    \caption{Molecular structures of (a) $p$-terphenyl, (b) pentacene, and (c) deuterated-pentacene.}
    \label{fig:molecular-structure}
\end{figure}

Figure~\ref{fig:molecular-structure} shows the molecular structures of $p$-terphenyl, pentacene, and deuterated-pentacene.

\newpage

\section{Derivation of the optically initialized electron spin populations in NV$^-$ center}
\label{sec:derivation-initial-population}
While intersystem crossing (ISC) to a triplet state distributes populations over the eigenstates $\{|X\rangle, |Y\rangle, |Z\rangle \}$ of the ZFS Hamiltonian $\mathcal{H}_{\mathrm{ZFS}}=D\left[S_{z}^{2} - S(S+1) \right] + E (S_{x}^{2} - S_{y}^{2})$ \textit{alone} (with no static field), many experiments are subject to an external magnetic field where the system's eigenstates $\{ |\psi_{1}\rangle, |\psi_{0}\rangle, |\psi_{-1}\rangle \}$ are represented by field-dependent linear combinations of the ZFS eigenstates. Since the high-field eigenstates $|\psi_{m}\rangle$ $(m=0,\pm 1)$ and the zero-field eigenstates $\{|X\rangle, |Y\rangle, |Z\rangle \}$ are related through unitary transformation, populations over the former can be calculated given the populations over the latter, and vice versa. Our aim here is to estimate the populations $\rho_{X}, \rho_{Y}$, and $\rho_{Z}$ over NV$^{-}$'s ZFS eigenstates from the populations $\rho_{m}$ over the eigenstates $|\psi_{m} \rangle$ in a given high field, and then to extract the populations in the \textit{different} magnetic field that the overtone EPR experiments were performed under the common condition for pulsed laser excitation. Note that this section's discussion is specific to NV$^{-}$ in which the populations over the ZFS eigenstates of the \textit{ground} triplet state upon pulsed laser irradiation depend on how many cycles of photo-exitation and ISC processes take place during the pulsed laser irradiation. This is in contrast to pentacene, where the populations over the ZFS eigenstates of the metastable excited triplet state are determined by a single ISC process. Also note that, since $|X \rangle$ and $|Y \rangle$ are degenerate states for $E=0$ (as is the case for NV$^{-}$), they are equally populated, i.e.,
\begin{align}
  \rho_{X} = \rho_{Y} \equiv p, 
\end{align}
so that, from $\rho_{X}+\rho_{Y}+\rho_{Z}=1$,
\begin{align}
  \rho_{Z} = 1 - 2p.
\end{align}

In our previous work (Section S1 of Ref.~\cite{miyanishi2021room}), the electron spin polarization $P_{0,-1}$ of NV$^-$ between the states $|\psi_{0} \rangle$ and $|\psi_{-1}\rangle$ in a magnetic field of 0.36~T at 11.6~GHz was determined to be 0.16 in nanodiamonds using the same experimental setting as used in the present work.
That is,
\begin{align}
    P_{0,-1} = \left|{\frac{\rho_{0}-\rho_{-1}}{\rho_{0}+\rho_{-1}}}\right| = 0.16
    \label{eq:P^H_0-1}
\end{align}
in 0.36~T and at 11.6~GHz. This combination of the field and the frequency hits the resonance in those crystallites where the major principal axis ($Z$) of the ZFS tensor is normal to the direction of the magnetic field $\bm{B}$.
With $\bm{B} \parallel X$, the zero-field and high-field eigenstates are related through
\begin{align}
    |\psi_{1} \rangle &= \cos \left(\frac{\alpha}{2}\right) |Y \rangle + \sin\left(\frac{\alpha}{2}\right)|Z\rangle, \\
    |\psi_{0}\rangle &= |X \rangle, \\
    |\psi_{-1}\rangle &= -\sin \left(\frac{\alpha}{2}\right) |Y\rangle + \cos\left(\frac{\alpha}{2}\right)|Z \rangle, 
\end{align}
where $\alpha = \tan^{-1} (2\omega_{e}/D) = \tan^{-1}(2\gamma B/D)$ and
\begin{align}
    \cos\left( \frac{\alpha}{2}\right) &= \frac{1}{\sqrt{2}} \left[ 
    1 + \frac{1}{\sqrt{1+\left(\frac{2\gamma B}{D}\right)^{2}}}
    \right]^{\frac{1}{2}}, \\
    \sin\left( \frac{\alpha}{2}\right) &= \frac{1}{\sqrt{2}} \left[ 
    1 - \frac{1}{\sqrt{1+\left(\frac{2\gamma B}{D}\right)^{2}}}
    \right]^{\frac{1}{2}}.
\end{align}
It follows that
\begin{align}
    \rho_{1} &= \rho_{Y}\cos^{2}\left(\frac{\alpha}{2}\right) + \rho_{Z} \sin^{2}\left(\frac{\alpha}{2}\right)
            = p \cos^{2}\left(\frac{\alpha}{2}\right) + (1-2p)\sin^{2}\left(\frac{\alpha}{2}\right), \\
    \rho_{0} &= \rho_{X} = p, \\
    \rho_{-1} &= \rho_{Y}\sin^{2}\left(\frac{\alpha}{2}\right) + \rho_{Z} \cos^{2}\left(\frac{\alpha}{2}\right) 
           = p \sin^{2}\left(\frac{\alpha}{2}\right) + (1-2p) \cos^{2}\left(\frac{\alpha}{2}\right). 
\end{align}
With $\gamma B = 2\pi\cdot 11.6$~GHz and $D=2\pi\cdot 2.87$~GHz, $\cos^{2}(\alpha/2) \approx 0.561$ and $\sin^{2}(\alpha/2)\approx 0.439$.
By plugging the above equations into Eq.~(\ref{eq:P^H_0-1}), we obtain
\begin{align}
  0.16 &= 
  \frac{p \left( -1+\sin^{2}\left( \frac{\alpha}{2}\right)\right) + (1-2p)\cos^{2}\left( \frac{\alpha}{2}\right)}{p\left( 1+\sin^{2}\left( \frac{\alpha}{2}\right)\right) + (1-2p)\cos^{2}\left( \frac{\alpha}{2}\right)} \\
  & = \frac{(1-3p)\cos^{2}\left(\frac{\alpha}{2}\right)}
  {(1-p)+(-1+3p)\sin^{2}\left(\frac{\alpha}{2}\right)},
\end{align}
whence $p=0.27$ and 
\begin{align}
  2p &=0.54, \\
  (1-2p)& =0.46.
\end{align}

\newpage

\section{Calculation of overtone polarization and powder average}
Now that the populations $\rho_{X}$, $\rho_{Y}$, and $\rho_{Z}$ over the ZFS eigenstates $\left\{ |X\rangle, |Y\rangle, |Z\rangle \right\}$ have been determined, we are ready to numerically calculate the populations $\rho_{m}$ over the eigenstates $| \psi_{m}\rangle$ for arbitrary strength $B$ and orientation $(\beta,\gamma)$ of the magnetic field $\bm{B}$. 
With the basis set
\begin{align}
    |X\rangle &= \frac{1}{\sqrt{2}} \left( -|1\rangle + |-1\rangle \right), \\
    |Y\rangle &= \frac{1}{\sqrt{2}} \left( |1\rangle + |-1\rangle \right), \\
    |Z \rangle &= |0\rangle,
\end{align}
where $|\pm 1\rangle$ and $|0\rangle$ are the eigenstates of $S_{z}$ with spin 1, the total Hamiltonian $H_{\mathrm{static,e}} = H_{\mathrm{Zeeman,e}}+H_{\mathrm{ZFS}}$ is represented by a matrix
\begin{align}
  \left[ 
    \begin{array}{ccc}
      \frac{1}{3}D-E & -B\cos\beta & iB\sin\beta\sin\gamma \\
      -B\cos\beta & \frac{1}{3}D+E & B\sin\beta\cos\gamma \\
      -iB\sin\beta\sin\gamma & B\sin\beta\cos\gamma & -\frac{2}{3}D
    \end{array}
  \right].
\end{align}
Note that $E=0$ for NV$^{-}$ and $|E|=2\pi\cdot 53.5$~MHz for pentacene.
By numerically calculating the unitary matrix $V$ that diagonalizes $H_{\mathrm{static,e}}$ such that
\begin{align}
    V^{\dagger} H_{\mathrm{static,e}} V = \mathrm{diag}(\omega_{1},\omega_{0},\omega_{-1}),
\end{align}
or 
\begin{align}
    H_{\mathrm{static,e}} V = V \cdot \mathrm{diag}(\omega_{1},\omega_{0},\omega_{-1}),
\end{align}
we obtain
\begin{align}
  | \psi_{1}(\beta,\gamma) \rangle &= V_{11}|X\rangle + V_{21} |Y\rangle + V_{31} |Z \rangle, \\
  | \psi_{0}(\beta,\gamma) \rangle &= V_{12}|X\rangle + V_{22} |Y\rangle + V_{32} |Z \rangle, \\
  | \psi_{-1}(\beta,\gamma) &= V_{13}|X\rangle + V_{23} |Y\rangle + V_{33} |Z \rangle,
\end{align}
and
\begin{align}
  \rho_{1}(\beta,\gamma) = |V_{11}|^{2} \rho_{X} + |V_{21}|^{2} \rho_{X} + |V_{31}|^{2} \rho_{Z}, \\
  \rho_{0}(\beta,\gamma) = |V_{12}|^{2} \rho_{X} + |V_{22}|^{2} \rho_{X} + |V_{32}|^{2} \rho_{Z}, \\
  \rho_{-1}(\beta,\gamma) = |V_{13}|^{2} \rho_{X} + |V_{23}|^{2} \rho_{X} + |V_{33}|^{2} \rho_{Z}.
\end{align}
Numerical values of the anisotropic overtone polarization $P_{1,-1}(\beta,\gamma)$, now given by
\begin{align}
    P_{1,-1}(\beta,\gamma) = \frac{\rho_{1}(\beta,\gamma)-\rho_{-1}(\beta,\gamma)}{\rho_{1}(\beta,\gamma)+\rho_{-1}(\beta,\gamma)},
\end{align}
are surface-plotted in Fig.~4 for pentacene and NV$^{-}$. The average values $\overline{P_{1,-1}}$,
\begin{align} 
    \overline{P_{1,-1}} = \frac{1}{4\pi} \int^{2\pi}_{0} d\gamma \int^{\pi}_{0} \sin\beta P_{1,-1}(\beta,\gamma)
\end{align}
were 0.071 for pentacene and 0.074 for NV$^{-}$ in a magnetic field $B$ of 0.207~T and 0.196~T, respectively.

\newpage

\section{Hyperpolarization experiment using pentacene-$d_{14}$-doped $p$-terphenyl}
\begin{figure}[h]
    \includegraphics[width=130mm]{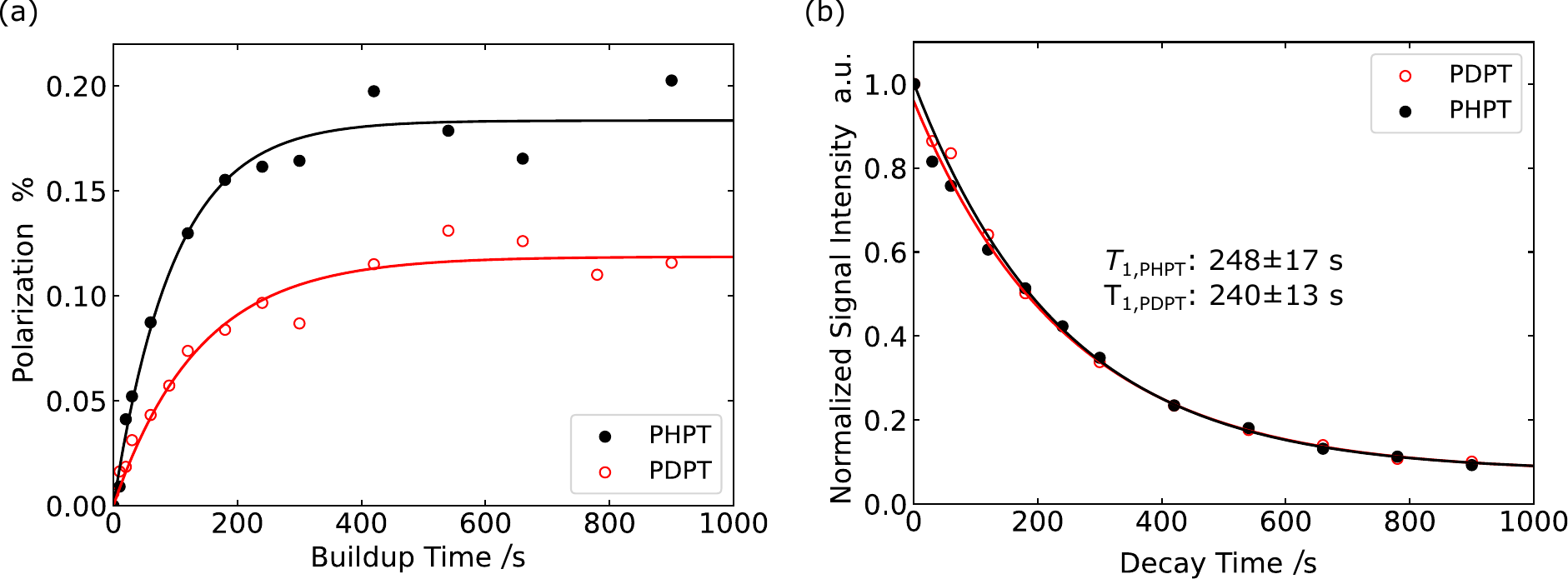}
    \caption{(a) Buildup curves of $^1$H polarization for PHPT (black) and PDPT (red) with ISE repetition rates of 50~Hz. (b) Relaxation curves for the $^1$H NMR signal of PHPT (black) and PDPT (red).}
    \label{fig:PandT_PDPTandPHPT}
\end{figure}

We performed hyperpolarization of the $^1$H spins of 0.57~mg of pentacene-$d_{14}$-doped $p$-terphenyl~(PDPT) sample using the ISE pulse sequence shown in Fig.~5(a).
The same parameters for the ISE pulse sequence for PHPT were used.
The polarization and relaxation time were obtained by curve fitting of the data points in Fig.~\ref{fig:PandT_PDPTandPHPT} to be 0.119$\pm$0.004\% and 240$\pm$13~s.

\begin{table*}[htb]
  \begin{tabular}{cccc} \hline \hline
    Sample  & PHPT (\DeltaMs) & PDPT (\DeltaMs) & PHPT (\DeltaMsone)~\cite{kagawa2023triplet} \\ \hline
    $B_{0}$~/~T & \multicolumn{2}{c}{0.207} & 0.39  \\ \hline
    $P$ /~\% & 0.183$\pm$0.005 & 0.119$\pm$0.004 & 0.75$\pm$0.02   \\ \hline
    $T_1$ ~/~s & 248$\pm$17 & 240$\pm$13 & 137$\pm$9 \\ \hline 
    \hline
  \end{tabular}
    \caption{Magnetic field strength, obtained polarization, and relaxation time.}
    \label{Table:ParametersForSampleAndOptimisedSequence}
\end{table*}

%